\documentclass[alpha-refs]{wiley-article}

\usepackage{color}
\usepackage{ulem}
\usepackage{amsmath, amssymb, mathrsfs, dsfont, yfonts}
\usepackage{bm}  
\usepackage{bbm}

\usepackage{mathtools}
\usepackage{amsfonts}
\usepackage{xparse}

\usepackage{cases}

\usepackage{nicematrix}
\usepackage[table]{xcolor}
\usepackage{transparent}
\usepackage{multirow}
\usepackage{multicol}
\usepackage{booktabs}
\usepackage{float}
\usepackage{caption}
\usepackage{subcaption}

\usepackage[utf8]{inputenc}
\RequirePackage[T1]{fontenc}

\definecolor{airforceblue}{rgb}{0.36, 0.54, 0.66} 
\definecolor{lapislazuli}{rgb}{0.15, 0.38, 0.61}
\definecolor{oxfordblue}{rgb}{0.0, 0.13, 0.28}
\usepackage{hyperref}
\hypersetup{backref=true,       
                    pagebackref=true,               
                    hyperindex=true,                
                    colorlinks=true,                
                    breaklinks=true,                
                    urlcolor= black,                
                    linkcolor= oxfordblue,                
                    bookmarks=true,                 
                    bookmarksopen=false,
                    filecolor=black,
                    citecolor=oxfordblue,
                    linkbordercolor=oxfordblue
}
\AtBeginDocument{\hypersetup{pdfborder={0 0 1}}}

\usepackage{url}            
\usepackage{booktabs}       
\usepackage{amsfonts}       
\usepackage{nicefrac}       
\usepackage{microtype}      
\usepackage{lipsum}
\usepackage{fancyhdr}       
\usepackage{graphicx}       
\graphicspath{{media/}}     
\usepackage{times}
\usepackage{url}
\usepackage{latexsym}
\usepackage{afterpage}
\def\bSig\mathbf{\Sigma}

\definecolor{mycolor1}{rgb}{0.00000,0.44700,0.74100}%
\definecolor{mycolor2}{rgb}{0.85000,0.32500,0.09800}%
\definecolor{mycolor3}{rgb}{0.92900,0.69400,0.12500}%
\definecolor{mycolor4}{rgb}{0.49400,0.18400,0.55600}%
\definecolor{mycolor5}{rgb}{0.46600,0.67400,0.18800}%
\definecolor{mycolor6}{rgb}{0.30100,0.74500,0.93300}%

\tikzset{
  dot to dot/.style={
    every node/.style={inner sep=1.5pt, draw, circle}
  }
}
\NewDocumentCommand \dotToDot { s } {
  \IfBooleanTF{#1}{
    \tikzset{left dot/.style={fill}, right dot/.style={}}    
  }{
    \tikzset{left dot/.style={}, right dot/.style={fill}}
  }
  \mathrel{
    \begin{tikzpicture}[
        baseline={(0, -.5ex)},
        dot to dot
      ] 
      \node[left dot]  (left)              {};
      \node[right dot] (right) at (3em, 0) {};
      \draw (left) -- (right);
    \end{tikzpicture}
  }
}

\usepackage[parfill]{parskip}
\papertype{Original Article}
\paperfield{Journal Section}

\title{Bayesian model-based outlier detection in network meta-analysis}

\author[1\authfn{1}]{Silvia Metelli}
\author[2\authfn{2}]{Dimitris Mavridis}
\author[1,3\authfn{3}]{Perrine Cr\'equit}
\author[1]{Anna Chaimani}

\affil[1]{Inserm Research Center of Epidemiology and Statistics, Universit\'{e} Paris Cit\'{e}, France}
\affil[2]{Department of Primary Education, University of Ioannina, Greece}
\affil[3]{Direction de la recherche Clinique, H\^{o}pital Foch, Suresnes, France}
\corraddress{Silvia Metelli PhD, Inserm Research Center of Epidemiology and Statistics, Universit\'{e} Paris Cit\'{e}, France}
\corremail{silvia.metelli@u-paris.fr}

\presentadd[\authfn{2}]{Inserm Research Center of Epidemiology and Statistics, Universit\'{e} Paris Cit\'{e}, Paris, 75004, France}

\fundinginfo{This project has received funding from the European Union's Horizon 2020 research and innovation programme under the Marie Sk{\l}odowska-Curie grant agreement No 101031840.}

\runningauthor{Metelli et al.}

\begin{document}

\maketitle

\begin{abstract}
\setlength{\parskip}{0em}
In network meta-analysis, some of the collected studies may deviate markedly from the others, for example having very unusual effect sizes. These deviating studies can be regarded as outlying with respect to the rest of the network and can be influential on the pooled results. Thus, it could be inappropriate to synthesise those studies without further investigation. In this paper, we propose two Bayesian methods to detect outliers in a network meta-analysis via: (a) a mean-shifted outlier model and (b), posterior predictive $p$-values constructed from ad-hoc discrepancy measures. The former method uses Bayes factors to formally test each study against outliers while the latter provides a score of outlyingness for each study in the network, allowing to numerically quantify the uncertainty associated with being outlier. Furthermore, we present a simple method based on informative priors as part of the network meta-analysis model to down-weight the detected outliers. We conduct extensive simulations to evaluate the effectiveness of the proposed methodology while comparing it to some alternative outlier detection tools. Two case studies are then used to demonstrate our methods in practice. 
\keywords{outlying studies, indirect treatment effects, Bayes factors,  posterior predictive checking, down-weighting}
\end{abstract}

\section{Introduction}
\label{sec:intro}

{\color{black} In medical statistics, meta-analyses and network meta-analyses (NMAs) \citep{lum2002,lu2004} have become crucial tools to quantitatively pool results from independent studies and assess treatment efficacy and cost-effectiveness. In pairwise meta-analysis, only two treatments at the time can be compared, while network meta-analysis allows for the simultaneous comparison of multiple ($\ge 3$) treatments, forming a so-called \textit{network} of treatments.} By integrating into a single model direct and indirect evidence across trials, network meta-analysis has the potential to provide a more precise, global estimate of the relative effect of any pair of treatments included in the network. 
To avoid misleading conclusions and provide valuable information for clinical decisions, the  network needs to be carefully screened looking for studies with markedly different or extreme effect sizes, namely outlying studies. Outliers may occur for many different reasons, including very small sample sizes or study-specific effect sizes whose distribution depart from the conventional normal curve (e.g. heavy-tailed or skewed distribution of the effect sizes). Such studies can substantially influence and alter the conclusions of the analysis and need proper investigation. 

Whilst many different issues of network meta-analysis methodology, such as inconsistency and heterogeneity, have received large attention in the literature, outlying studies - although intrinsically related to the presence of inconsistency and heterogeneity in the network - have not been widely studied. To date, no specific guidelines exist for how these studies should be treated in the general context of evidence synthesis. Several outlier detection methods have been recently developed for pairwise meta-analysis \citep{viech2010, Gumedze2011, Zhao2017, Mavridis2017} but little work has been done to extend the methods to network meta-analysis. Moreover, most of the available techniques are based on useful yet heuristic diagnostics measures such as studentised residuals or the Cook's distance while only a few rely on probabilistic model-based approaches. Among these, a frequentist `variance shift' outlier model has been proposed for univariate meta-analysis \citep{Gumedze2011} while two `mean-shift' models have been later developed for a bivariate model for diagnostic test accuracy (DTA) meta-analyses and subsequently for a full multivariate model for network meta-analysis \citep{Negeri2020, Noma2020}. In both cases, the methodology made use of a frequentist likelihood ratio test (LRT) as a test statistic for assessing whether each included study was outlying and the parametric bootstrap approach to approximate the sampling distribution of the observed LRT statistic. Bayesian approaches are attractive in network meta-analysis \citep{Dias2018} and have the advantage of using the exact likelihood for the data (i.e. binomial for binary data) rather than relying on normal approximations. However, outlier detection in the Bayesian framework has not been sufficiently explored, with exception of one method for pairwise meta-analysis of DTAs \citep{Matsushima2018} and one introducing  a Bayesian $p$-value for network meta-analysis but mainly focusing on arm-based models for continuous outcomes \citep{Zhang2015}. 

A comprehensive assessment against outlyingness should not merely focus on the statistical detection of extreme effect measures (or variances) of the studies included; rather, it should try to understand the causes behind it through a careful appraisal of the characteristics of each included study. A related question then arises about how these studies should be treated while ensuring that the validity and robustness of the synthesis process is maintained. The debate was initially centered around whether or not outliers should be removed from the analysis \citep{Hedges1985}. Conducting sensitivity analyses with and without outliers to monitor the changes in the summary effects is surely useful, but clinicians might still not reach consensus about which scenario should be used for their final clinical decisions. Therefore, more tailored strategies for treating outliers are necessary. For example, methods have been proposed in pairwise meta-analysis for building heterogeneity measures which are minimally affected by the presence of outliers \citep{Lin2017} or down-weighting the apparent outlying studies without removing them \citep{Gumedze2011}. In network meta-analysis, this also ensures that the connectivity of the network is maintained.

In this paper, we suggest to employ a two-step procedure: first, a probabilistic outlier detection model is used to quantify outlying behaviour and then, the studies associated with high probability of being outliers can undergo down-weighting, if appropriate. As a first step towards this, we propose an intuitive Bayesian mean-shift model that detects deviating studies within the network using Bayes factors; then we seek to complement Bayes factor detection with Bayesian model checking, which allows to better quantify the associated uncertainty for each study to be outlier. Specifically, we propose the use of posterior predictive $p$-values under ad-hoc discrepancy measures, which  are well-suited to capture local deviations in the model. As a second step, informative beta priors are conveniently incorporated into the network meta-analysis model to down-weight the outliers identified. The performance of our methods is assessed and compared using both simulated and real data.

{\color{black} The rest of the paper is structured as follows. Section \ref{sec:data} describes two examples of real networks of treatments while in Section \ref{sec:standarNMA} we briefly introduce the most commonly used random effects network meta-analysis model. In Section \ref{sec:outliers}, we describe our proposed approaches: first, a mean-shifted model with Bayes factors and then, posterior predictive checks with ad-hoc discrepancy measures; while the down-weighting scheme is described in Section \ref{sec:down}. In Section \ref{sec:simul}, we perform an extensive simulation study and in Section \ref{sec:appli} we present an application to the two real networks previously introduced. Finally, we conclude with a discussion in Section \ref{sec:concl}.}

{\color{black}
\section{Exemplar data}
\label{sec:data}
}

\noindent We introduce two real data sets, each forming a network of treatments, which we later use to demonstrate our methods. The first example is a network of treatments for non-small cell lung cancer (NSCLC) and the second is a smaller network of non-pharmacological interventions for smoking cessation. Non-small cell lung cancer represents approximately 85\% of all lung cancer cases, and most patients have wild-type or unknown status for epidermal growth factor receptor (EGFR) which often leads to a diagnosis of advanced-stage disease. According to specific eligibility criteria, patients with advanced-stage diagnosis might receive second-line treatments instead of palliative care. Despite the American Society of Clinical Oncology recommends two cytotoxic drugs and two EGFR-tyrosine kinase inhibitors \citep{oncol2015}, many new treatments have been recently approved by the US Food and Drug Administration (FDA) and over the years, more than forty treatments have been assessed in randomised trials for second-line treatment of advanced NSCLC \citep{crequit2016}. Clearly, simultaneously comparing the relative efficacy and safety of all available treatments in a network can better assist clinical decision-making. \citealt{crequit2017} conducted an extensive systematic review and NMA for second-line treatments of advanced NSCLC (involving a total of 39,388 patients), forming a network of $N=112$ randomised controlled trials (RCTs) comparing $62$ different treatments, many of which informed by one or very few studies only. This makes it a good candidate to suspect the presence of trials with outlying results. As a second example, we use a well-known network of $N=24$ RCTs investigating four different counselling programs to aid smoking cessation (involving a total of 16,737 participants, \citealt{Hasselblad}, where the four counselling interventions are defined as self-help, individual counselling, group counselling, and no contact. This network is mainly used in this article for comparison purposes, as it has been recently tested for the presence of outliers in a network meta-analysis application \citep{Petropoulou2021}. 

\begin{figure}[bt]
\begin{center}
	\vspace*{0.7cm}
	\includegraphics[scale=0.065]{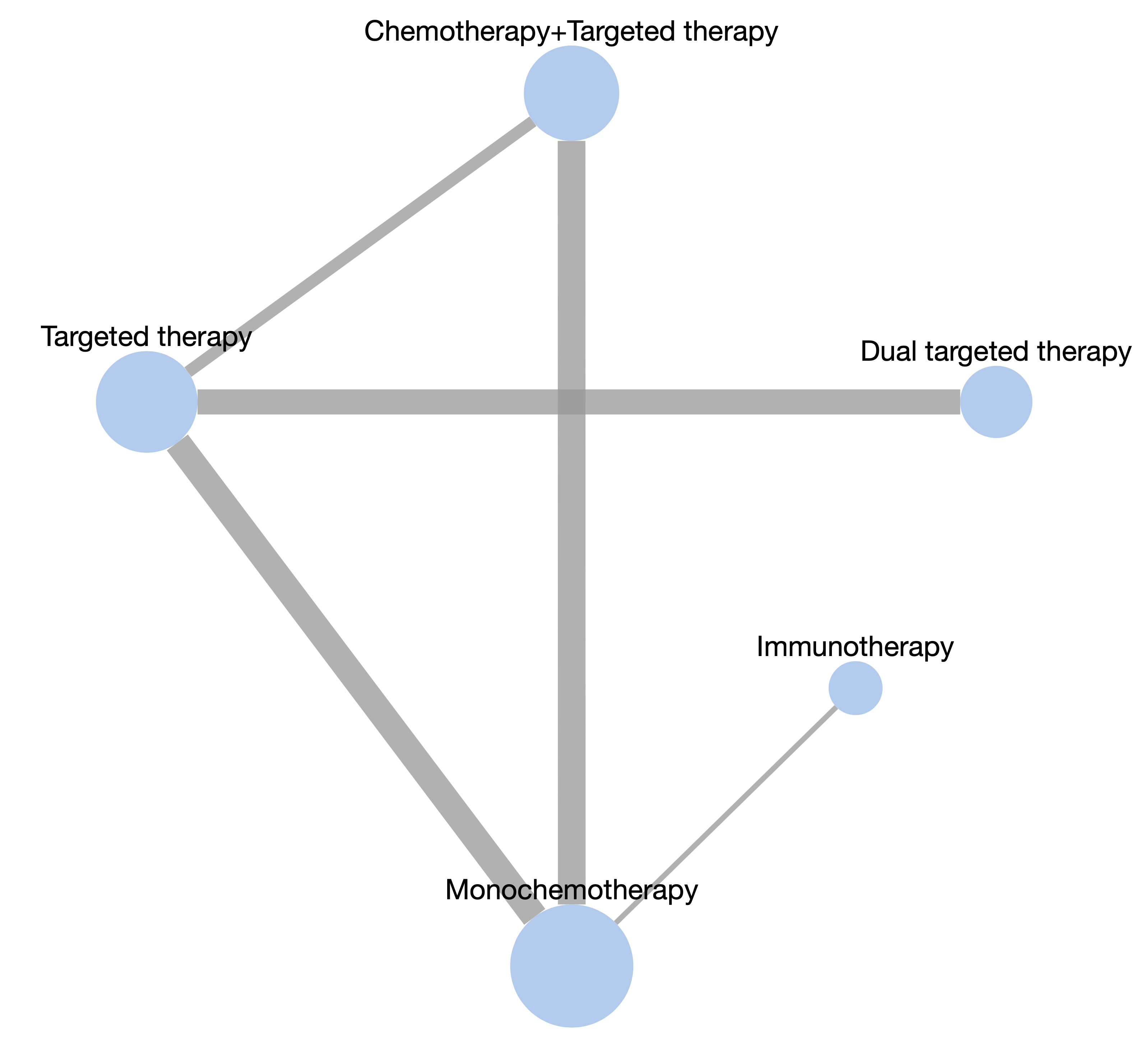} \hspace{0.15cm}
	\includegraphics[scale=0.062]{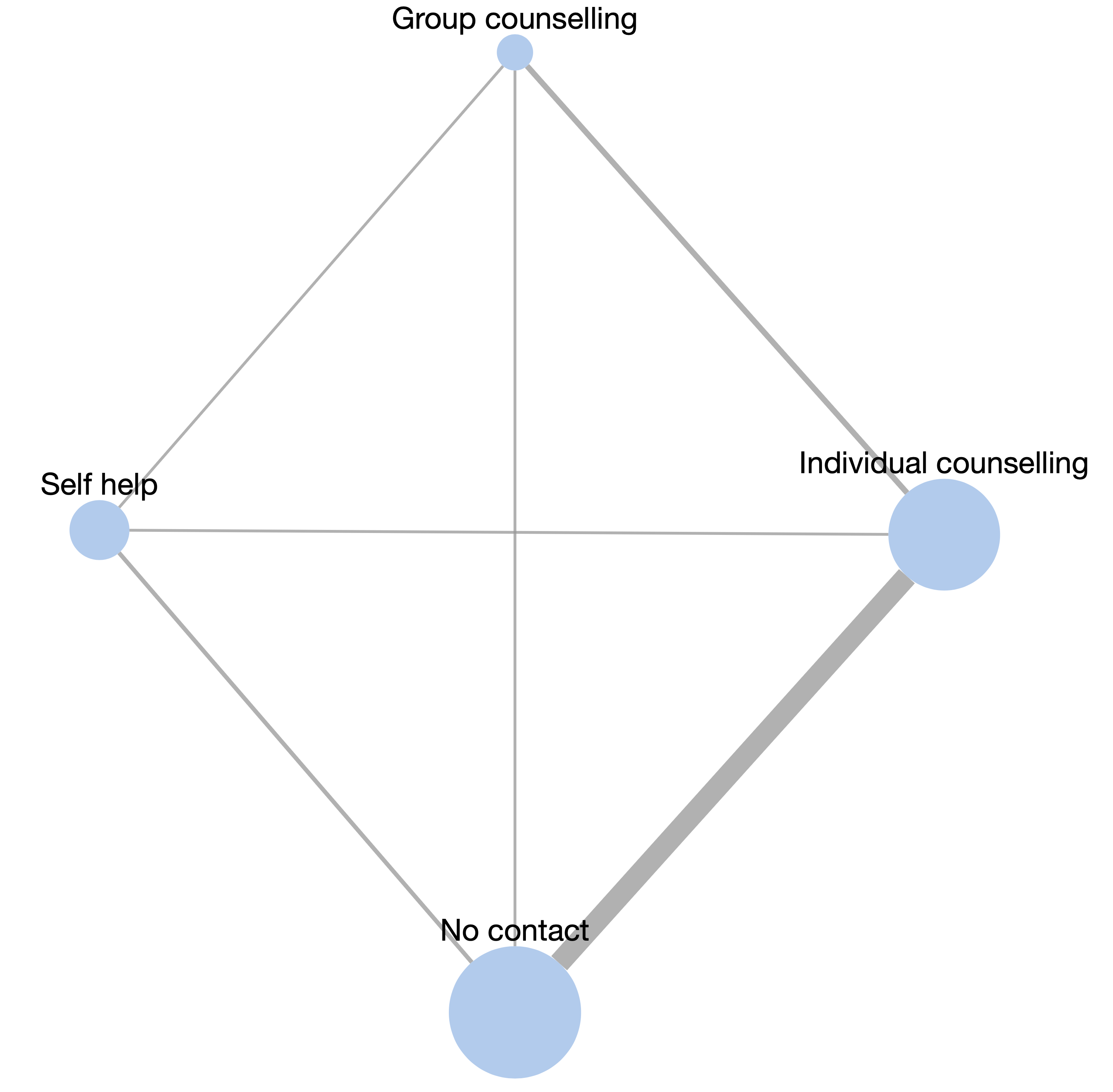}
		\vspace*{0.15cm}

\end{center}
\caption{A network of second-line treatment classes for non-small cell lung cancer (left) and a network of counselling interventions for smoking cessation (right). Node size is proportional to the number of individuals randomised, while edge size is proportional to the number of studies available for that comparison.}	
\label{fig:fig_netplots}
\end{figure}

In Figure \ref{fig:fig_netplots}, we show the network geometries for both data sets. For NSCLC data, each treatment was further grouped to one out of five treatment classes (Dual Targeted Therapy, Chemotherapy plus Targeted Therapy, Immunotherapy, Monochemotherapy and Targeted Therapy) to facilitate the visualisation of the available evidence. The full NSCLC network at treatment level can be found in the Supplementary material.\\[2pt]

\section{Network meta-analysis random effects model}
\label{sec:standarNMA}

{\color{black}
\subsection{Basic elements and notation}
\label{subsec:nma}
}

\noindent Network meta-analyses expand the scope of more conventional pairwise meta-analyses to simultaneously compare multiple treatments in a connected network of evidence, where the information of the relative treatment effects (e.g. log odds ratios) is pooled across multiple studies \citep{lu2006}. More specifically, consider a collection of studies $i=1\dots,N$, where each study $i$ only compares a subset $\mathcal{K}_i$ of the full set of $\{1, \dots, K\}$ treatments. Let $k_i=|\mathcal{K}_i|$ be the cardinality of $\mathcal{K}_i$, then in most NMAs, $k_i$ is 2 (``two-arm study'') or 3 and rarely we have studies with four or more arms (``multi-arm studies''). In the following, we focus on NMAs with a binary outcome (e.g. death, no death), so for each study $i$ we have data $\mathcal{D}=\{\left(r_{ik},n_{ik}\right):i=1,\ldots,N;k\in \mathcal{K}_i\}$, where $r_{ik}$ is the number of observed events and $n_{ik}$ the total number of participants for the $k{\textsuperscript{th}}$ treatment in the $i{{\textsuperscript{th}}}$ study. The corresponding probability of the event will be denoted by $\pi_{ik}$.

\subsection{Standard model}
 {\color{black}{
 In each study $i$, let one treatment be seen as the baseline treatment, $b_i$ (simply denoted as $b$ in the following for convenience). Without loss of generality, the baseline treatment can be considered a reference (e.g. placebo) against which each other treatment $k \in \{1, \dots, K\}/\{b\}$ is compared. Then, the commonly used random-effects network meta-analysis for the binomial data can be written as}}
\begin{align}
\label{eq:model}
\begin{split}
& r_{ik} \sim \mbox{Binomial}\left(n_{ik}, \pi_{ik}\right),	\quad i=1,\dots,N, k \in \mathcal{K}_i\\
& {\mbox{logit}\left(\pi_{ik}\right)=}\mu_i+\theta_{bk}+\delta_{i,bk}, \quad k \neq b,
\end{split}
\end{align}
 {\color{black}{where $\mu_i$ represents the log odds of the baseline treatment $b$ in each study $i$, since for $k=b$ the logit expression in \eqref{eq:model} simply reduces to ${\mbox{logit}\left(\pi_{ik}\right)=}\mu_i$.}} This parameter is generally considered a nuisance while the main interest lies in the mean relative effect $\theta_{bk}$. Likewise, for continuous outcome data or log odds and risk ratios we can formulate the same network meta-analysis model using a normal likelihood with the identity link function instead of the logit one. To be identifiable, the model requires an arbitrary reference treatment whose effect is set to zero. Here, we choose reference $b=1$ so that $\boldsymbol{\theta}=\left(\theta_{12},\theta_{13},\ldots,\theta_{1K}\right)^{T}$ is a vector of treatment effects relative to the reference treatment, which are called the basic parameters. Then, assuming statistical consistency, i.e. agreement between direct and indirect evidence, we have $\theta_{hk}=\ \theta_{1k}-\theta_{1h}$ for every treatment pairs $\left(h,k\right)\in\{1,\ldots,K\}$.  {\color{black}{In words, when we have both direct and indirect evidence for a particular comparison, then consistency holds in the data if no discrepancy exists in the treatment effects obtained under both types of evidence.}} All other relative effects can be obtained as linear combinations of the basic parameters in $\boldsymbol{\theta}$. 

Study-specific heterogeneity is captured by the random effects $\delta_{i,bk}$, which represent the relative effects between treatment $k$ and $b$ for the $i^{th}$ study. We assume exchangeability of the $\delta_{i,bk}$ so that the NMA model provides estimates for the $\theta_{bk}$'s, and the between-study heterogeneity variance of the random effects $\tau_{bk}^2$. The specific distributional assumptions made on $\delta_{i,bk}$ are discussed separately below. {\color{black} Suppose $\boldsymbol{\delta}_i=\left(\delta_{i,12},\delta_{i,13},\ldots,\delta_{i,1K}\right)^{T}$ is the vector of study-specific relative effects of treatment $k$ versus $b=1$.} Then, ${\ \boldsymbol{\delta}}_i\in\mathbb{R}^{k_i-1}$ is assumed to follow a multivariate normal distribution, 
\begin{equation}
\boldsymbol{\delta}_i\sim N\left(\mathbf{0},\boldsymbol{\Psi}_i^2\right).	
\end{equation}
\noindent Following \citet{Higgins_white} and \citet{lum2002}, we assume throughout the paper a common heterogeneity, i.e. $\tau_{bk}=\tau$, for all comparisons and the $(k_i-1)\times(k_i-1)$ matrix $\boldsymbol{\Psi}_i^2$  to be homogeneous and symmetric with $\tau^2$ elements on the diagonal representing treatment-specific variances and $\tau^2/2$ elements off-diagonal, representing between-study covariance.  {\color{black} This, along with the consistency equation, ensures that in each study $i$, with treatment pair $(h,k)$, $Var(\delta_{i,hk})=Var(\delta_{i,1h})+Var(\delta_{i,1k})-2Cov(\delta_{i,1h},\delta_{i,1k}) \Leftrightarrow \tau^2 = 2\tau^2-2Cov(\delta_{i,1h},\delta_{i,1k}) \Leftrightarrow  Cov(\delta_{i,1h},\delta_{i,1k})=\tau^2/2$.}

Finally, the observed data $\mathcal{D}$ are described by the following likelihood function:
\begin{equation}
P\left(\mathcal{D}\middle|\boldsymbol{\mu}, \boldsymbol{\theta}, \tau^2\right)=\prod_{i=1}^{N} { \prod_{k\in \mathcal{K}_i} {\small \binom{n_{ik}}{r_{ik}}} \left[\mbox{logit}^{-1}(\pi_{ik}) \right]^{r_{ik}} \left[1-\mbox{logit}^{-1}(\pi_{ik}) \right]^{n_{ik}-r_{ik}}} 
\label{eq:modellik}
\end{equation}
{\color{black}{
\noindent where $\boldsymbol{\mu}=\left(\mu_1,\dots, \mu_N\right)$ is the vector of study-specific baseline parameters (intercepts), while $\boldsymbol{\theta}$ and $\tau^2$ are the parameters of primary interest. To estimate parameters, NMA models are often rely on maximum likelihood, but the hierarchical structure of random-effect components typically requires numerical optimisation methods or restricted maximum likelihood (REML) techniques. Here, we take a Bayesian approach, and so parameters of interest are assigned independent prior distributions, $P(\boldsymbol{\mu})$, $P(\boldsymbol{\theta})$ and $P(\tau^2)$. Posterior inference is then conducted on the joint posterior distribution of parameters, 
\begin{equation}
P\left(\boldsymbol{\mu}, \boldsymbol{\theta}, \tau^2 \middle| \mathcal{D}\right) \propto P\left(\mathcal{D}\middle|\boldsymbol{\mu}, \boldsymbol{\theta}, \tau^2\right)P(\boldsymbol{\mu})P(\boldsymbol{\theta})P(\tau^2).
\label{eq:posterior}
\end{equation}
As exact inference on this posterior distribution is not analytically tractable, Markov chain Monte Carlo (MCMC) simulation is used to perform posterior inference. 
}}

\section{Outlier detection}
\label{sec:outliers}

\subsection{Mean-shift model with Bayes factor tests}

We define outliers in a network meta-analysis as studies with `shifted' effect sizes and we propose a mean-shifted model to identify such studies.  
This model assumes `shifted' location parameters for the effect sizes of study $i$, meaning that the underlying relative effect in the $i$-th study may diverge from those of the other studies. In practice, this means that model \ref{eq:model} is replaced by 
\begin{align}
\label{eq:model2}
\begin{split}
& r_{ik} \sim \mbox{Binomial}\left(n_{ik}, \pi_{ik}\right),	\quad i=1,\dots,N, k \in \mathcal{K}_i\\
& {\mbox{logit}\left(\tilde \pi_{ik}\right)=}\mu_i+\theta_{bk}+\eta_{bk}+\delta_{i,bk}, \quad k \neq b.
\end{split}
\end{align}

In this case, the likelihood function contains an additional parameter vector: 
\begin{equation}
P\left(\mathcal{D}\middle|\boldsymbol{\mu}, \boldsymbol{\theta},\boldsymbol{\eta},\tau^2\right)= \prod_{i=1}^{N}  \prod_{k\in \mathcal{K}_i}  \binom{n_{ik}}{r_{ik}} \left[{\mbox{logit}}^{-1}(\tilde \pi_{ik}) \right]^{r_{ik}}  \left[1-{\mbox{logit}}^{-1}(\tilde \pi_{ik}) \right]^{n_{ik}-r_{ik}}
\label{eq:modellikshift}
\end{equation}
\noindent where $\boldsymbol{\eta}=\left(\eta_{12},\eta_{13}, \ldots,\eta_{1K}\right)^T$ is a vector of mean-shift location parameters, {\color{black} i.e. the grand mean parameters of study $i$ may deviate from the grand mean of the other studies, and this implicitly means that we are assuming a mean-shifted model for the random-effect of each study $i$.} Then, the outlier detection problem can be cast as follows: if the mean-shifted model with non-zero shift factors is more plausible than the ordinary network meta-analysis model, then the $i$-th study can be seen a potential outlier. This corresponds to testing the following hypothesis for each study $i$: 
\begin{equation}
 H_0: \ {\eta}_{bk}=0\ \ \ \ vs.\ \ \ \ H_1: \eta_{bk}\neq0, \ \ \ \ \\ \quad b\neq k,\ \forall\ k\in\{1,\ldots,K\}/\{1\}	
\label{eq:test}
\end{equation}
In a Bayesian hypothesis testing context, the test above can be formally assessed through Bayes factors (BFs). Suppose model 0 $(H_0)$ is the standard model and model 1 $(H_1)$  is the mean-shift outlier model, then the Bayes factor would take the following form:
\begin{equation}
BF_{1:0}=\frac{P\left(\mathcal{D}\middle| H_1\right)}{P\left(\mathcal{D}\middle| H_0\right)}=\frac{\int_{ \Theta_1}{P\left(\mathcal{D}\middle|\boldsymbol{\theta}_1 \right)P_1(\boldsymbol{\theta}_1)d{\boldsymbol{\theta}_1}}}{\int_{ \Theta_0}{P\left(\mathcal{D}\middle|\boldsymbol{\theta}_0 \right)P_0(\boldsymbol{\theta}_0)d{\boldsymbol{\theta}_0}}},	
\end{equation}

\noindent with $\boldsymbol{\theta}_1=(\boldsymbol{\mu}, \boldsymbol{\theta}, \boldsymbol{\eta}, \tau^2) \in \Theta_1$, $\boldsymbol{\theta}_0=(\boldsymbol{\mu}, \boldsymbol{\theta}, \tau^2) \in \Theta_0$, $P_{0}(\boldsymbol{\theta}_0)$ and $P_{1}(\boldsymbol{\theta}_1)$ being the prior distributions for the parameters of interest. The Bayes factor can be interpreted as an updating factor of prior beliefs, and represents how likely the data were predicted by $H_1$ compared to $H_0$. This also provides a fair comparison between two models of different parameter dimension, since the Bayesian paradigm embodies a natural penalty against overfitting, i.e. the Occam's razor principle. 

{\color{black}
This model can be seen as the Bayesian counterpart of the location-shift model introduced by \citealt{Noma2020}, where potential outliers were searched via bootstrap-adjusted Likelihood Ratio tests. Our model has the additional flexibility provided by prior information, which is crucial in NMA outlier-detection as we often encounter small or sparse networks informed by a few studies only. Furthermore, being ratios of probabilities, Bayes factors also give an indication about the size of the evidence. Indeed, they represent the relative probability assigned to the observed data under each of the two hypotheses, and so they not only provide evidence in favour of outlyingness, $H_1$, as the classical hypothesis testing, but also in favour of $H_0$. }

\subsection{Posterior predictive model checking}
\label{sec:pppc}

An alternative possibility for Bayesian model-based outlier detection is posterior predictive checking \citep{Meng1994, Gelman1996}, which is a commonly used tool for the identification of divergent observations of Bayesian models. The idea is to construct a discrepancy measure which captures deviation between the observed data and the posterior predictive distribution of the assumed model, which is for us the standard random effects network meta-analysis model. First, we take the posterior predictive distribution by simulating replicated data from the fitted model. Then we compare the replicated to the observed data to look for systematic discrepancies that will show us whether the observed data could have been plausible under the hypothesised model. {\color{black} The discrepancy measure $f$ is often taken to be the omnibus $\chi^2$ measure proposed by \citealt{Gelman1996}. This approach was followed in \citealt{Zhang2015} to construct a `Bayesian $p$-value'. However the method is primarily built for arm-based NMA models and continuous data where absolute treatment effects are assumed exchangeable while here we focus on contrast-based modelling where exchangeability is assumed on relative treatment effects, as introduced in Section \ref{sec:standarNMA}. In simulations (see Supplementary material), we found that Gelman's discrepancy performs poorly in the present context. Omnibus discrepancy measures are useful but provide less power with respect to measures designed to test specific features of the data (e.g. extremeness), suggesting the need for a discrepancy measure more capable to detect local deviations in the model. } Thus, we propose two different choices for $f$: first, we make use of the single log-likelihood contribution of each study $i$ and then we leverage the Stahel-Donoho outlyingness (SDO) measure \citep{Stahel1981, Donoho1982} to construct an `outlyingness score'. For each study $i$ with arm data $D_{i,k}=(r_{ik},n_{i,k})$, the two discrepancy measures are respectively given by
\begin{align}
f_i^{\textsuperscript{L}}=\sum_{k\in K_i}\mbox{log}P\left(D_{i,k}\middle|\boldsymbol{\theta}_0\right) \label{eq:D1}, \\
f_i^{\textsuperscript{SDO}}=\sum_{k\in K_i}\frac{\left|x_{i,k}-\mbox{med}\left(\boldsymbol{x}\right)\right|}{\mbox{MAD}\left(\boldsymbol{x}\right)} \label{eq:D2},	
\end{align}
{\color{black} where $x_{ik}=r_{ik}/n_{ik}$, $\mbox{med}\left(\boldsymbol{x}\right)$ is the median and $\mbox{MAD}\left(\boldsymbol{x}\right)=med_i(|x_i-med_j(x_j)|)$ is the median absolute deviation of the observed proportions $x_{ik}: i=1,\dots,N; k\in K_i$.} The first proposal is somewhat related to the omnibus $\chi^2$ measure but captures different aspects of the relationship between data structure and the parameters and avoids producing extremely small values in presence of studies with small variances, while the second is specifically aimed at detecting asymmetry in the data. Note that the first measure depends both on data and model parameters while the second depends on the data only.

The values of the discrepancy measure for the observed data are compared to values of the posterior predictive distribution: large differences indicate lack of fit. Specifically, we use posterior predictive $p$-values, which calculate a tail-area probability given that the assumed model is true, and so quantify the extremeness of the observed value rather than offering a strict accept-reject decision rule as in standard hypothesis testing. An extreme $p$-value implies that the observed data would be unlikely to occur in replications of the data if the model was true and so, may represent an outlier. Here, posterior predictive $p$-values quantify the uncertainty associated with each study in the network by measuring departure of each study from the assumed model. For each study $i$ let $\mathcal{D}_i=\{ (r_{i,k},n_{i,k}): k\in K_i\}$, then the \textit{posterior predictive $p$-value} is as follows:
\begin{equation}
{p}_{f_i} \equiv P\left\{f_i\left(\mathcal{D}_i^\ast\boldsymbol|\boldsymbol{\theta}_0\right)\geq f_i\left(\mathcal{D}_i\middle|\boldsymbol{\theta}_0\right)\middle|\mathcal{D}\right\} =
\int{P\left\{f_i\left(\mathcal{D}_i^\ast\middle|\boldsymbol{\theta}_0\right)\geq f_i\left(\mathcal{D}_i\middle|\boldsymbol{\theta}_0\right)\middle|\boldsymbol{\theta}_0\right\}}P\left(\boldsymbol{\theta}_0\middle|\mathcal{D}\right)d\boldsymbol{\theta}_0,
\label{eq:pval}
\end{equation}

where $\mathcal{D}$ is the observed data, $\mathcal{D}^\ast$ a hypothetical replicated data set generated from the model predictive distribution, and $P\left\{\cdot\middle|\mathcal{D}\right\}$ the joint posterior distribution of $\left(\boldsymbol{\theta}_0,\mathcal{D}^\ast\right)$ given $\mathcal{D}$. 
{\color{black} This can be easily estimated from the MCMC samples as 
\begin{equation}
{p}_{f_i} = \frac{1}{S} \sum_{s=1}^{S} \mathbbm{1}  \left\{f_i\left(\mathcal{D}_i^\ast\middle|\boldsymbol{\theta}_0(s)\right)\geq f_i\left(\mathcal{D}_i\middle|\boldsymbol{\theta}_0(s)\right)\right\}
\label{eq:pval2}
\end{equation}
where $S$ is the number of MCMC simulations and $\boldsymbol{\theta}_0(s)$ the simulated parameter values at step $s$.} Plugging-in the discrepancies of \eqref{eq:D1} and \eqref{eq:D2} into the $p$-value in \eqref{eq:pval2} we obtain our proposed posterior predictive $p$-values under the two different discrepancies, {\color{black} which we respectively denote $p_{{\textsubscript{L}}}$ and $p_{{\textsubscript{SDO}}}$}.

%

\section{Down-weighting outliers}
\label{sec:down}
Statistical detection of outlying effects in a network meta-analysis should always be complemented by an accurate investigation of the causes underlying the observed outlyingness. Before taking any decision, investigators should carefully check the characteristics of all included studies looking for possible explanations, such as systematic differences that modify the observed effect and produce extreme results. In particular, characteristics of the trial design, conduct, participants, interventions and outcomes should be explicitly assessed. Placing more stringent inclusion criteria in the systematic review may not always capture differences when they are subtle, and thus, a thorough assessment of the nature and reliability of the data is always necessary. 

When no clear causes are identified, it is possible to construct systems to down-weight the effect of outlying studies towards the overall network estimates, which seems a more reasonable choice compared to removing outlying studies \textit{tout-court} from the analysis.{\color{black} Indeed, the latter approach comes at the risk of disconnecting the network graph and this would prevent the whole NMA analysis. The risk is particularly high for sparse networks, where some comparisons might be informed by only one study.

We propose a computationally simple scheme that consists of two-stages: first, we screen the studies looking for outliers using the two methods described in the previous sections, i.e. for each study $i=1, \dots, N$ we calculate Bayes Factors and posterior predictive $p$-values.  If a study is associated with either a Bayes Factor above the chosen outlying threshold and/or a posterior predictive $p$-value below the significance threshold, further investigation is conducted. Then, if down-weighting is deemed appropriate, a second stage of analysis is performed where informative power priors \citep{Ibrahim2000} are used to automatically raise the likelihood of each outlying study $j$ to a power strictly between 0 and 1, to reduce its impact on the overall results. Here, that power represents a down-weighting factor $w_j \in\left(0,1\right)$. At the second stage, the joint posterior in \eqref{eq:posterior} is modified to 
\begin{equation}
P\left(\boldsymbol{\mu}, \boldsymbol{\theta}, \tau^2 \middle| \mathcal{D}\right) \propto P\left(\mathcal{D}^{{o}}\middle|\boldsymbol{\mu}, \boldsymbol{\theta}, \tau^2\right)P\left(\mathcal{D}^{\overline{o}}\middle|\boldsymbol{\mu}, \boldsymbol{\theta}, \tau^2\right)P(\boldsymbol{\mu})P(\boldsymbol{\theta})P(\tau^2),
\label{eq:posterior_dw}
\end{equation}
where $\mathcal{D}^{{o}}= \{(r_{jk},n_{jk}): j=1,\dots, N_{o}; k \in \mathcal{K}_{j} \}$ is the sub-set containing data for outlying studies, with size $N_{o}$. Analogously, we can define $\mathcal{D}^{\bar o}=\mathcal{D} \setminus \mathcal{D}^{{o}}$ as the set of data for the remaining $N-N_{o}$ non-outlying studies. In expression \eqref{eq:posterior_dw}, $P\left(\mathcal{D}^{{\overline{o}}}\middle|\boldsymbol{\mu}, \boldsymbol{\theta}, \tau^2\right)$ is defined as in \eqref{eq:modellik} with the only difference of using the restricted set of data $\mathcal{D}^{\overline{o}}$ while
\begin{equation}
P\left(\mathcal{D}^{o}\middle|\boldsymbol{\mu}, \boldsymbol{\theta}, \tau^2\right)  = \prod_{j=1}^{N_o} \prod_{k\in \mathcal{K}_j}   \left[  \binom{n_{jk}}{r_{jk}} \left[\mbox{logit}^{-1}(\pi_{jk}) \right]^{r_{jk}}   
 \left[1-\mbox{logit}^{-1}(\pi_{jk}) \right]^{n_{jk}-r_{jk}} \right]^{w_j},
\label{eq:lik_powerprior}
\end{equation}

\noindent where $\mbox{logit}(\pi_{jk})=\mu_j+\theta_{bk}+\delta_{j,bk}$ for each outlying study $j$.} As per Bayesian approach, the down-weighting factors $w_j$ are treated themselves as random variables and hence assigned their own prior distributions. {\color{black} We choose informative beta priors ${\ w}_{j\ } \sim \mbox{Beta}\left(a_j,b_j\right)$, so that the hyperparameters $a_j$ and $b_j$ can be specified to reflect how unusual the outlying study $j$ appears to be:} they can be centered at values $\le 0.5$ if we seek to apply a severe down-weighting - for example if there is additional external evidence supporting our hypothesis - or conversely, centered at values $\ge 0.5$ if we seek to apply a moderate down-weight - for example when being more uncertain about whether the study is an actual outlier or not. Examples of beta distributions reflecting different prior belief scenarios can be found in the Supplementary material. Ideally, external opinion should be used to elicit the beta distribution incorporating information from experts about their level of trust of suspicious effect sizes or studies. This approach would be particularly beneficial in the presence of so-called ``mega-trials'' with large discrepancies between fixed and random effects pooled estimates.  In presence of heterogeneity, a random effect model would indeed give a large weight to small studies: if appropriate, our scheme could down-weight such studies according to expert information.

\section{Simulation study}
\label{sec:simul}

We conducted a simulation study to assess the performance of our outlier detection tools on binary outcome data. We constructed four different network geometries and we analyzed a number of different scenarios, varying the amount of heterogeneity and number of outliers included in the network. For each scenario, we simulated $r=1000$ data sets for two- and multi-arm trials by drawing study-specific treatment effects $\boldsymbol{\theta}$ and covariance matrix $\boldsymbol{\Psi}_i^2$, as defined in Section \ref{sec:standarNMA}. In all scenarios we sampled two MCMC chains, with $50000$ iterations and a burn-in period of size $10000$. {\color{black} Vague normal priors, $N(0,1000)$, were used for the fixed effect and for each basic parameter and location-shift parameter. A vague uniform distribution, $U(0,5)$, was used for the heterogeneity $\tau^2$} and a beta prior centred around $0.5$, i.e. $w_i \sim \mbox{Beta}\left(3,3\right)$, was used for the down-weighting factors, to reflect a moderate down-weighting.

{\color{black}
The proposed methods have been then compared to similar approaches available in the literature, namely the Likelihood Ratio approach proposed in \citealt{Noma2020} where bootstrapped outlier $p$-values are approximated, and the Bayesian $p$-value proposed in \citealt{Zhang2015}. In addition, we have compared the methods to two cross-validatory leave-one-out alternatives, namely a recently developed Forward Search (FS) algorithm \citep{Petropoulou2021} and the Conditional Predictive Ordinate (CPO) diagnostics \citep{Gelfand1995}}. The former monitors several diagnostic measures in a forward fashion, i.e. starting from a basic `outlier-free' set of studies and sequentially adding the remaining studies, while the latter is a Bayesian diagnostic to detect surprising observations. CPO values where estimated via integrated nested Laplace approximations (INLA) \citep{Rue2009, Held2010, Held2015}, which is an available alternative to MCMC to estimate Bayesian models. Specifically, INLA offers a convenient and fast way to perform cross-validatory posterior predictive checking in a Bayesian framework without re-running the model during the backward search. {\color{black} Extreme values of the diagnostics measures monitored during the FS search (here Cook's distance larger than 1) as well as large CPO values (typically extreme values above 70) may indicate outlying observations \citep{Ntzoufras}.}

\subsection{Simulation settings and data generation}

The number of studies per comparison was set to $10$ for all comparisons in an ideally balanced design and ranged from $1$ to $9$ to reflect values more often encountered in practice in three different unbalanced designs. 
The number of patients per trial arm was simulated from a uniform distribution $U\left(50,200\right)$ rounding to the closest integer. The number of studies per comparison was set to $10$ for all comparisons in an ideally balanced design and ranged from $1$ to $9$ to reflect values more often encountered in practice in three different unbalanced designs. The number of patients per trial arm was simulated from a uniform distribution $U\left(50,200\right)$ rounding to the closest integer. To calculate the probability of an event in each study treatment arm we first draw baseline risks, i.e. event rates for the reference (treatment 1), from a uniform distribution $\pi_{1,i}\sim U\left(0.4,0.6\right)$ and then back-calculate the probability of an event in each study treatment arm using both baseline risks, assuming an overall event risk of $0.5$. Then, the study arm-specific number of events was generated from a binomial distribution, using the probability of an event and the number of patients per trial-arm. We set the underlying true log odds ratios $\boldsymbol{\theta}^{true}$ of each treatment versus reference to be fixed at equal intervals between $0$ and $1$. Variances of the simulated log odds ratios are sampled from $U\left(s_{min}^2,s_{max}^2\right)$ with $\left(s_{min}^2,s_{max}^2\right)$ either $\left(2^2,\ {3.5}^2\right)$ or $\left({0.5}^2,\ 2^2\right)$ to represent different between-study variation. To study our detection power at varying heterogeneity, we analyse several between-study heterogeneity values chosen accordingly to the predictive distributions for heterogeneity estimated empirically by \citealt{Turner2012}, who elicit predictive distributions for heterogeneity expected in future meta-analyses. Different distributions are obtained for different settings defined by the type of outcome and intervention comparison. Specifically, we take $\tau^2\in\{0,\ 0.032,\ 0.096,\ 0.287\}$, which respectively correspond to no heterogeneity, first, second and third quartiles of the estimated distribution of heterogeneity (we choose the setting with subjective outcome and pharmacological vs. pharmacological intervention comparisons). Again, this choice of outcome and comparison was made to reflect common settings of published meta-analyses \citep{Nikolakopoulou2014}. {\color{black} Finally, we contaminate the data with $1$ or $3$ outlying log odds ratios, which are sampled from $N\left(\theta_{bk}\ \pm\ \ C,{\ \tau}^2\right)$ , where $C=2.5\sqrt{\left(s_{max}^2+\tau^2\right)}$ or $C=3\sqrt{\left(s_{max}^2+\tau^2\right)}$, corresponding respectively to less extreme and more extreme outliers. Number of events are then sampled accordingly from binomial distributions. Overall, we explored 32 different scenarios. A detailed summary of the different scenarios is reported in Table \ref{tab:scenarios}.  }

\begin{table}[!h]
\caption{ Summary of the different scenarios analysed in the simulation study.}
\centering
\resizebox{0.99\textwidth}{!}{%
\setlength{\tabcolsep}{10pt}
\setlength\extrarowheight{0.1pt}
\begin{NiceTabular}{ccc|cc}
\hline	
\rowcolor{white} \textbf{Network geometry} & \textbf{Design}  & \textbf{Scenario} & \textbf{Heterogeneity} & \textbf{Number of outliers} \\ \hline 
\multirow{8}{*}{\parbox[l]{7em}{ \hspace*{-1.95cm}
\includegraphics[width=2.3in, height=1.85in]{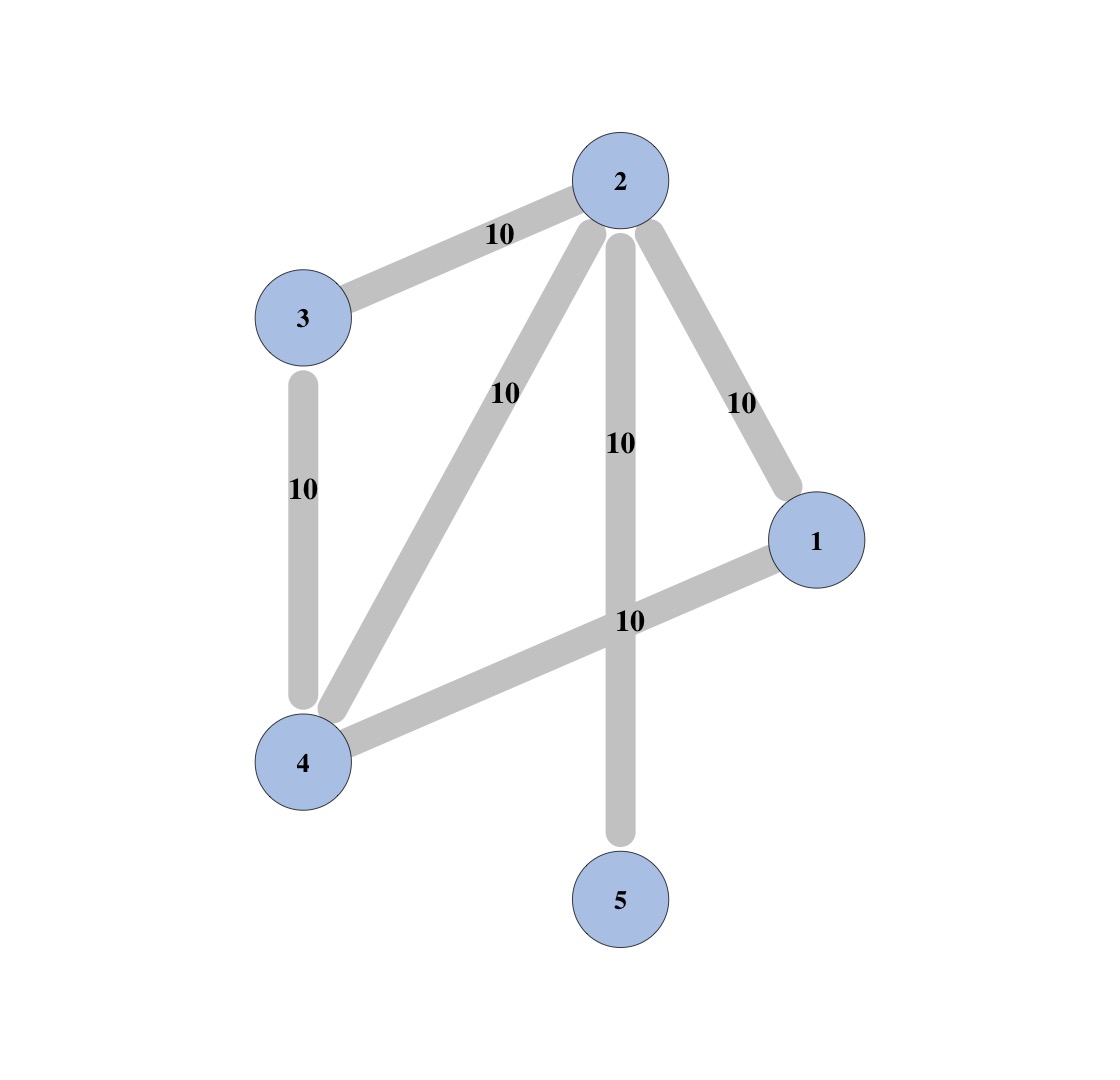}}}  &
 \multirow{8}{*}{{\begin{tabular}[c]{@{}>{\columncolor{black!5}}c@{}}Balanced\\ fairly-connected network \\{\small(100 studies in total)}\end{tabular}}}
&  1               	 	   & \cellcolor{black!0} 0                      & \cellcolor{black!0} 1                           \\
 & &	2					   & 0.032                  & 1                           \\
 &  &  3                         & \cellcolor{black!0} 0.096                  & \cellcolor{black!0} 1                           \\
 & & 4                         & 0.287                  & 1                           \\
 &  &5                  		   &\cellcolor{black!0}  0                      & \cellcolor{black!0} 3                           \\
 & & 6 						   & 0.032                  & 3                           \\
 &  &  7                         & \cellcolor{black!0} 0.096                  & \cellcolor{black!0} 3                           \\
 & & 8                         & 0.287                  & 3                           \\ \hline
 \multirow{8}{*}{\parbox[l]{7em}{ \hspace*{-1.95cm}
      \includegraphics[width=2.3in,height=1.85in]{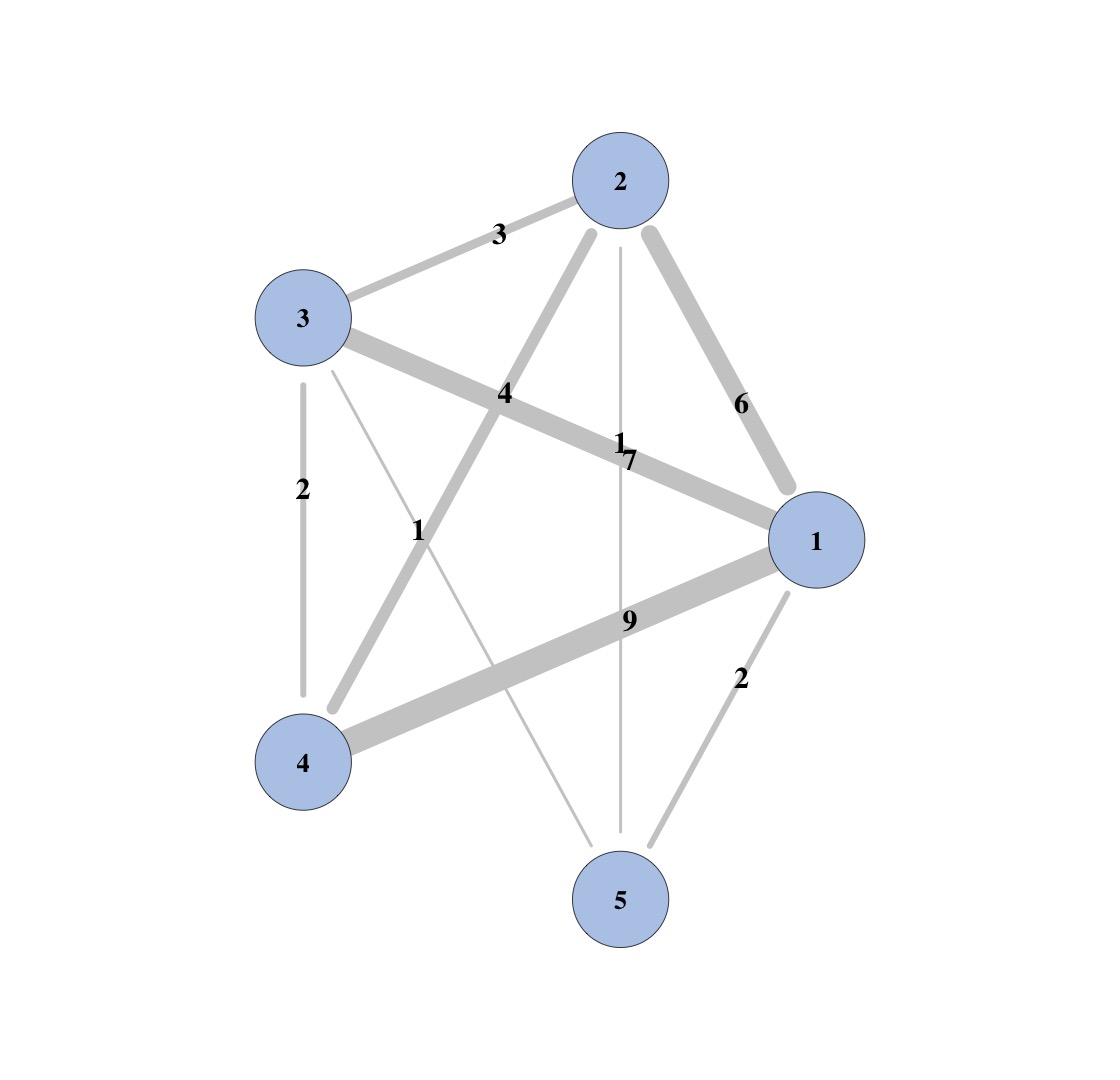}}}&
\multirow{8}{*}{{\begin{tabular}[c]{@{}>{\columncolor{black!5}}c@{}}Unbalanced\\ well-connected   network \\ {\small(35 studies in total)} \end{tabular}}} 
 \cellcolor{white} & 9                			& \cellcolor{black!0} 0                      & \cellcolor{black!0} 1                           \\
 && 10 							& 0.032                  & 1                           \\
 && 11                          & \cellcolor{black!0} 0.096                  & \cellcolor{black!0} 1                           \\
 && 12                          & 0.287                  & 1                           \\
 && 13                	      	& \cellcolor{black!0} 0                      & \cellcolor{black!0} 3                           \\
 && 14 						  	& 0.032                  & 3                           \\
 && 15                          & \cellcolor{black!0} 0.096                  & \cellcolor{black!0} 3                           \\
 && 16                          & 0.287                  & 3                           \\ \hline
  \multirow{8}{*}{\parbox[l]{7em}{ \hspace*{-1.95cm}
      \includegraphics[width=2.3in,height=1.85in]{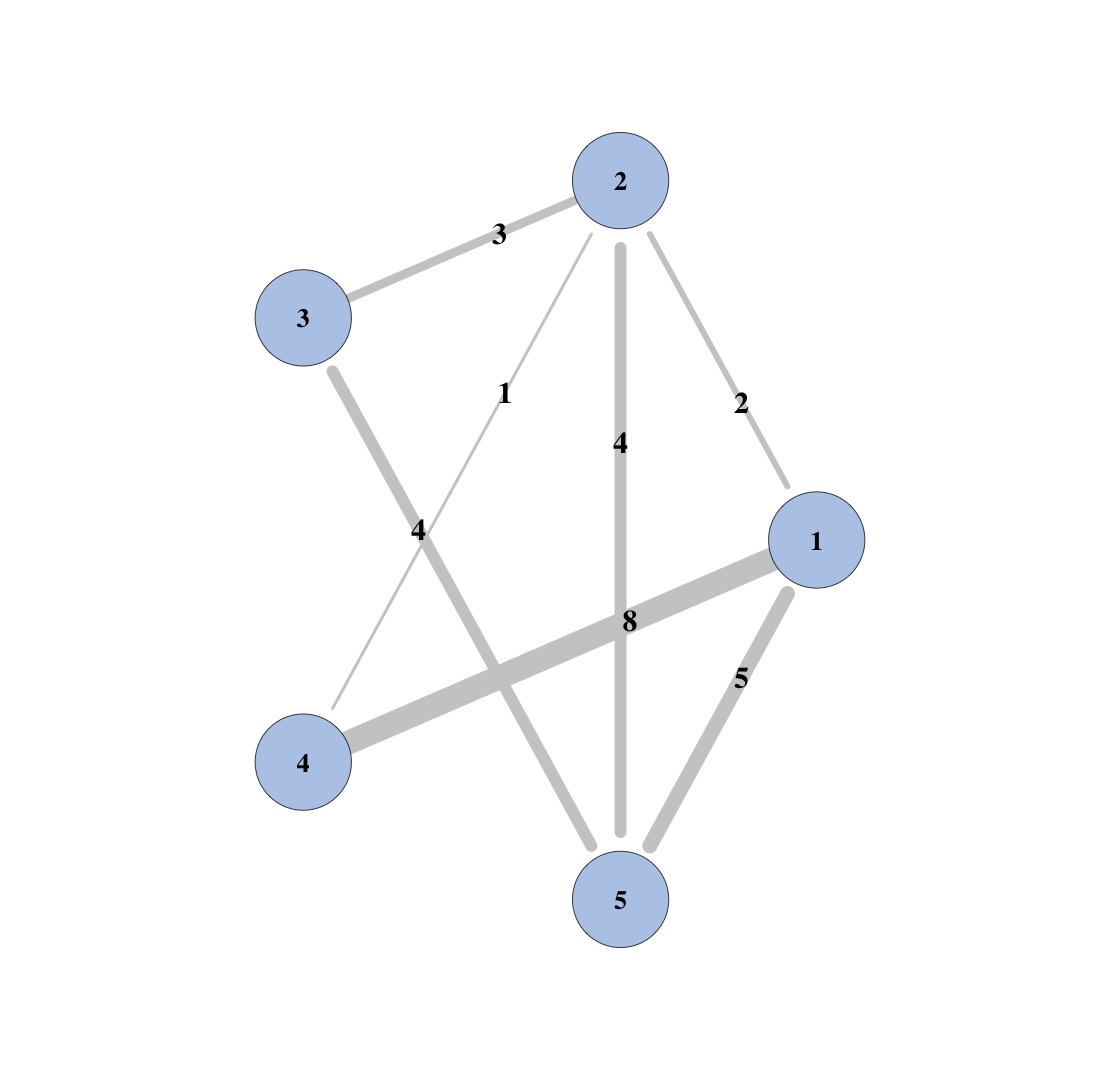}}}&
\multirow{8}{*}{{\begin{tabular}[c]{@{}>{\columncolor{black!5}}c@{}}Unbalanced\\ fairly-connected network \\ {\small(27 studies in total)}\end{tabular}}}         
 & 17                           & \cellcolor{black!0} 0                      & \cellcolor{black!0} 1                           \\
  && 18                           & 0.032                  &  1                           \\
  && 19                           &\cellcolor{black!0} 0.096                  & \cellcolor{black!0} 1                           \\
  && 20                           & 0.287                  & 1                           \\
  && 21                           & \cellcolor{black!0} 0                      & \cellcolor{black!0} 3                           \\
  && 22                           & 0.032                  & 3                           \\
  && 23                           & \cellcolor{black!0} 0.096                  & \cellcolor{black!0} 3                           \\
  && 24                           & 0.287                  & 3                           \\ \hline
   \multirow{8}{*}{\parbox[l]{7em}{ \hspace*{-1.95cm}
      \includegraphics[width=2.3in,height=1.85in]{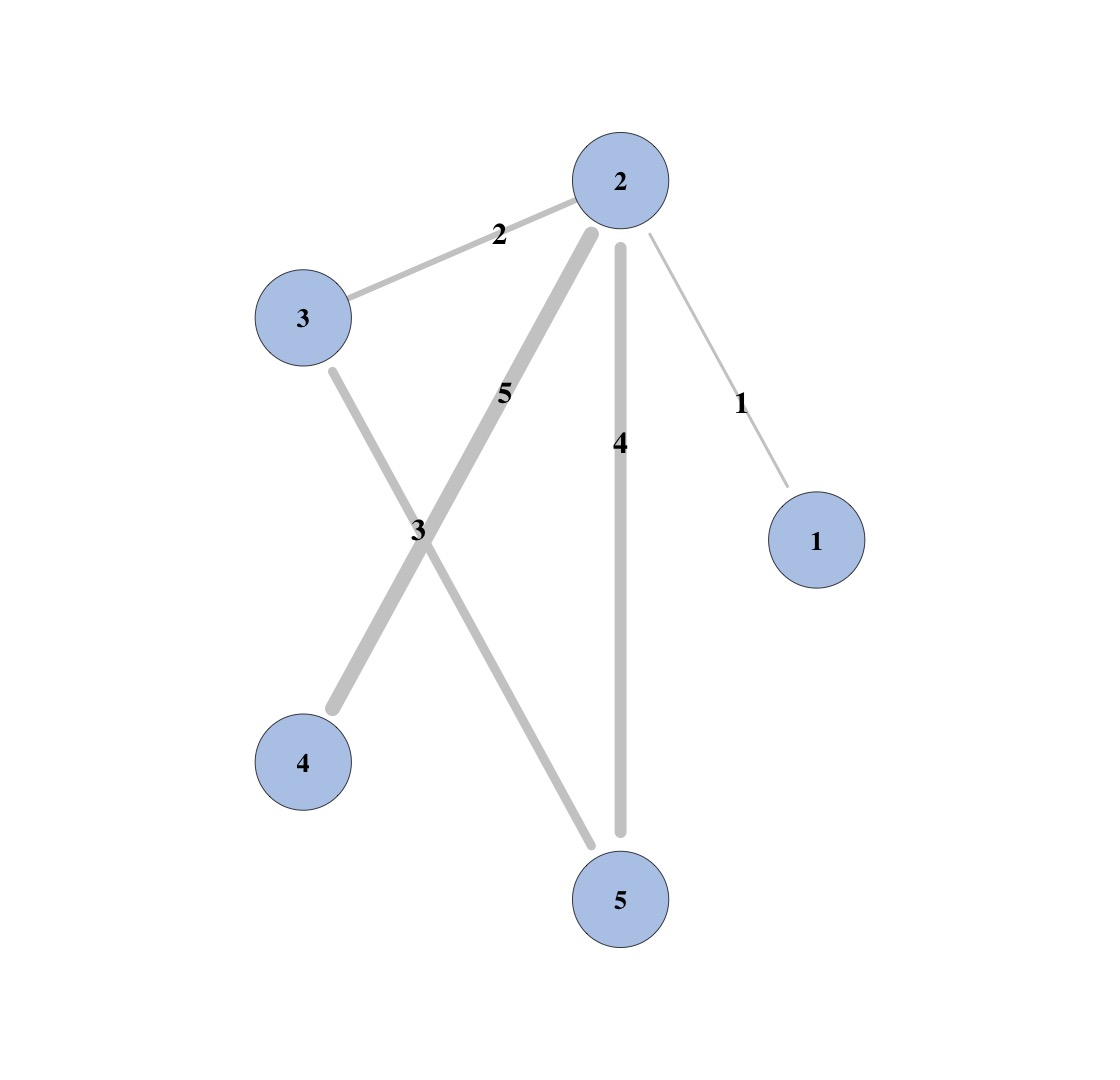}}}&
\multirow{8}{*}{{\begin{tabular}[c]{@{}>{\columncolor{black!5}}c@{}}Unbalanced\\ poorly-connected network \\{\small(15 studies in total)}\end{tabular}}}         
 & 25                           & \cellcolor{black!0} 0                      & \cellcolor{black!0} 1                           \\
  && 26                           & 0.032                  & 1                           \\
  && 27                           &\cellcolor{black!0}  0.096                  & \cellcolor{black!0} 1                           \\
  && 28                           & 0.287                  & 1                           \\
  && 29                           & \cellcolor{black!0} 0                      & \cellcolor{black!0} 3                           \\
  && 30                           & 0.032                  & 3                           \\
  && 31                           & \cellcolor{black!0} 0.096                  &\cellcolor{black!0}  3                           \\
  && 32                           & 0.287                  & 3                           \\ \hline
                      
\end{NiceTabular}
}
\label{tab:scenarios}
\end{table}

\subsection{Simulations results}

The main results of the simulation study are reported in Table \ref{tab1} and Table \ref{tab2}. The case of an unbalanced design with a fairly-connected network contaminated with either one or three outliers is chosen as a representative case often encountered in practice. To explore false-positive detections, we also included comparisons with the scenario where no outliers were induced. Additional results for all the included studies and remaining scenarios are reported in the Supplementary material. 
 
 {\color{black}
To assess our methods, we report the following performance measures.} First, we calculate mean Bayes Factors and mean posterior predictive $p$-values (under both discrepancy definitions). Then, we calculate and report the proportion of false-positive detections when no synthetic outliers are induced in the network and we compare with the false-positive rates obtained under competing methods. In addition, to assess the benefit of the down-weighting scheme we report the estimate relative bias for each treatment contrast, defined as  $({\hat{\theta}}{\textsuperscript{MC}}-\theta{\textsuperscript{true}})/\theta{\textsuperscript{true}}$, with ${\hat{\theta}}{\textsuperscript{MC}}$ Monte Carlo average of estimated effects. {\color{black} The evidence from Bayes factors is typically quantified as weak, moderate, strong or decisive through heuristic classification schemes  (see \citealt{kass1995} table in  Supplementary material). Following Kass and Raftery, we consider a study to show weak evidence of outlyingness if the Bayes Factor is above 3.2, decisive evidence if above 100, while as per standard convention we consider a study to show some evidence of outlyingness if the $p$-value is below 0.05. Similarly, details about the thresholds for detection used in the CPOs and FS algorithm are discussed in the Supplementary material.}

\begin{table}[!h]
\caption{Mean Bayes factors and mean posterior predictive $p$-values for the induced outliers of 1000 simulated data sets for the unbalanced design with a fairly connected network of 27 studies (scenarios 17-24 in Table \ref{tab:scenarios}). At varying scenarios, either a single outlier or three outliers (outlier 1, outlier 2, outlier 3) are induced in the network.  Here, BF: Bayes Factor test; LR: Likelihood Ratio test as in \citealt{Noma2020} (bootstrapped $p$-values reported); $\boldsymbol{p_{{\textsubscript{L}}}}$, $\boldsymbol{p_{{\textsubscript{SDO}}}}$ and $\boldsymbol{p_{{\textsubscript{G}}}}$ posterior predictive $p$-values under under likelihood-based discrepancy in \eqref{eq:D1}, Stahel-Donoho outlyingness discrepancy in \eqref{eq:D2}, and Gelman's Omnibus $\chi^2$ as in \citealt{Zhang2015}, CPO: conditional predictive ordinate values; FS: Forward search algorithm as in \citealt{Mavridis2017} (Cook's distance reported); $\tau^2$: heterogeneity (thresholds for detection: $BF>3.2$ for Bayes factors; $p<0.05$ for $p$-values, $CPO>70$ for conditional predictive ordinates and Cook's distance $>1$ for FS algorithm). }	

\centering
\resizebox{0.99\textwidth}{!}{%
\setlength{\tabcolsep}{12pt}
\setlength\extrarowheight{0.6pt}
\begin{NiceTabular}{c|c|ccc|cccc}
\hline	
\rowcolor{white} \textbf{ $\boldsymbol{\tau^2}$}  & \textbf{Induced outliers}   & \textbf{BF}     &$\boldsymbol{p_{{\textsubscript{L}}}}$ & $\boldsymbol{p_{{\textsubscript{SDO}}}}$ & \textbf{LR} &$\boldsymbol{p_{{\textsubscript{G}}}}$& \textbf{CPO} & \textbf{FS}   \\ \hline

 \multirow{4}{*}{ {\textbf{0}}}  & a single outlier & \cellcolor{black!0} 2063.3 &  \cellcolor{black!0} $<$0.001 &   \cellcolor{black!0} $<$0.001 & 0.01& $<$0.001 & 118 & 3.7  \\\cline{2-9}
     
   & outlier 1 & 511.1  & 0.01       & 0.05   & 0.01 &0.01 &  78 &  3.2       \\
   & outlier  2 &  \cellcolor{black!0}118.2  &  \cellcolor{black!0} 0.01             & \cellcolor{black!0} 0.001      & 0.02 & 0.01 & 115 &  2.1     \\
   & outlier 3  & 284.1  & 0.001      & 0.01    &  0.01 & 0.05 & 101 &   2.7     \\ \hline 
 
  \multirow{4}{*}{ {\textbf{0.032}} }    & a single outlier  &  \cellcolor{black!0}1540.1 & \cellcolor{black!0} 0.002  &  \cellcolor{black!0}$<$ 0.0001 & 0.01 & 0.01 & 132 & 3.0 \\ \cline{2-9}
     
   & outlier 1 & 287.1  & 0.05             & 0.01       &0.01 & 0.07 &68 & 2.8      \\
   & outlier 2         &  \cellcolor{black!0}452.1 &\cellcolor{black!0} 0.001    &  \cellcolor{black!0}0.001    & 0.01 & 0.02 & 80 &  2.2       \\
   & outlier 3 & 32.1   & 0.01             & 0.01   & 0.06& 0.06 & 99 &  2.9         \\ \hline 
 
 \multirow{4}{*}{    {\textbf{0.096}} }  & a single outlier &  \cellcolor{black!0}11.1   &  \cellcolor{black!0}0.05     &  \cellcolor{black!0}0.05       & 0.02 & 0.07  & 35 &  1.0     \\\cline{2-9} 
   
   &  outlier 1 & 9.1    & 0.06             & 0.06    &0.04 & 0.13 & 32 &   0.8       \\
   &  outlier 2  & \cellcolor{black!0} 2.7    & \cellcolor{black!0} 0.03             &  \cellcolor{black!0}0.05           & 0.05 & 0.05 & 46 & 0.6 \\
   & outlier 3    & 2.8    & 0.04             & 0.05     & 0.04 & 0.10 & 39 &  0.6       \\ \hline
  
  \multirow{4}{*}{{\textbf{0.287}} }  &  a single outlier   &  \cellcolor{black!0}3.5    & \cellcolor{black!0} 0.04     & \cellcolor{black!0} 0.08    & 0.06 & 0.07  & 30 & 0.3        \\\cline{2-9}  
                         
  & outlier 1 & 2.5    & 0.22             & 0.12       & 0.10 & 0.40 &10 &  0.6     \\
  & outlier 2  &  \cellcolor{black!0}1.3    &  \cellcolor{black!0}0.10  & \cellcolor{black!0}0.12      &0.10 & 0.25 & 22& 0.6  \\
  & outlier 3  & 0.98   & 0.17 & 0.15     & 0.12 & 0.31 & 29 &    0.5  	
\\ \bottomrule
\end{NiceTabular}
}
\label{tab1}
\end{table}

{\color{black}
Based on Table \ref{tab1}, both our Bayes Factor tests and posterior predictive $p$-values are able to detect the majority of the artificial outliers induced in the network (demonstrated by either large Bayes Factor and/or small $p$-value). In particular, our posterior predictive $p$-value based on the likelihood is able to identify some outliers in two highly heterogeneous scenarios where the Bayes factor tests fail (outlier 1 and outlier 2 for $\tau^2=0.096$, single outlier for $\tau^2=0.287$).} As we can see from Table \ref{tab1}, the detection performance is slightly higher when only one outlier is present in the network, and this might be due to the fact that multiple outliers can shift the overall network meta-analysis model estimates to an extent to which they are not anymore recognised as deviating. As might be expected, the detection becomes difficult at increasing heterogeneity. All the induced outliers are detected only when  heterogeneity is absent or low, $\tau^2=0$ and $\tau^2=0.032$; while only some outliers are detected for $\tau^2=0.096$ and very few when $\tau^2=0.287$ (see Supplementary material for all remaining scenarios). {\color{black}As expected, Bayes Factors and Likelihood Ratio tests have similar performance, while $p$-values based on Gelman's discrepancy \citep{Zhang2015} perform quite poorly in this context. This seems to be in line with results reported in \citet{Zhang2015}, which comment that their measure ``fails to uncover any outlyingness under the contrast-based framework, with all Bayesian $p$-values simply around 0.50". Cross-validatory CPO diagnostic and Forward Search based on Cook's distance perform quite well when low or moderate heterogeneity is present, but largely fail in highly heterogeneous scenarios. CPO diagnostics do not always discriminate well outliers from influential data, as points with high leverage may have small CPOs, independently of whether or not they are outliers. }

\begin{table}[bt]
\caption{
Proportion of false-positive detections, when no outlier is induced in the network, at varying simulation scenario. Here, BF: Bayes Factor tests; LR:  Likelihood Ratio test as in \citealt{Noma2020}; $\boldsymbol{p_{{\textsubscript{L}}}}$, $\boldsymbol{p_{{\textsubscript{SDO}}}}$ and $\boldsymbol{p_{{\textsubscript{G}}}}$ posterior predictive $p$-values under under likelihood-based discrepancy in \eqref{eq:D1}, Stahel-Donoho outlyingness discrepancy in \eqref{eq:D2}, and Gelman's Omnibus $\chi^2$ as in \citealt{Zhang2015}, CPO: conditional predictive ordinate values; FS: Forward search algorithm as in \citealt{Mavridis2017};  $\tau^2$: heterogeneity.
}
\centering
\resizebox{0.99\textwidth}{!}{%
\setlength{\tabcolsep}{10pt}
\setlength\extrarowheight{0.6pt}
\begin{NiceTabular}{cc|ccccccc}
\toprule
\rowcolor{white}  \textbf{Design} &$\boldsymbol{\tau^2}$  & \textbf{BF}     & \textbf{LR test} &$\boldsymbol{p_{{\textsubscript{L}}}}$ & $\boldsymbol{p_{{\textsubscript{SDO}}}}$  &$\boldsymbol{p_{{\textsubscript{G}}}}$& \textbf{CPO} & \textbf{FS}  \\	\hline

\multirow{4}{*}{{\begin{tabular}[c]{@{}>{\columncolor{black!5}}c@{}}Balanced\\ faily-connected   network \\ {\small(100 studies in total)} \end{tabular}}}  

& \textbf{0}       & 0       & 0    & 0     & 0     & 0 & 0     & 0   \\
& {\textbf{0.032}} & 0       & 0    & 0     & 0     & 0 & 0     & 0  \\
& {\textbf{0.096}} & 0       & 0    & 0     & 0     & 0.01 & 0.01  & 0.02   \\
& {\textbf{0.287}} & 0.1     & 0.1  & 0.1   & 0.01  & 0.02 & 0.03  & 0.02   \\  \hline 
\multirow{4}{*}{{\begin{tabular}[c]{@{}>{\columncolor{black!5}}c@{}}Unbalanced\\ well-connected   network \\ {\small(35 studies in total)} \end{tabular}}}  

& \textbf{0}       & 0   	  & 0      & 0      & 0  	   & 0     & 0.03 & 0     \\
& {\textbf{0.032}} & 0.03     & 0.03   & 0.03   & 0        & 0     & 0.03 & 0.03    \\
& {\textbf{0.096}} & 0.03     & 0.03   & 0.03   & 0  	   & 0.03  & 0.09 & 0.06    \\
& {\textbf{0.287}} & 0.06     & 0.06   & 0.03   & 0.03     & 0.06  & 0.11 & 0.06     \\ \hline 

\multirow{4}{*}{{\begin{tabular}[c]{@{}>{\columncolor{black!5}}c@{}}Unbalanced\\ fairly-connected network \\ {\small(27 studies in total)}\end{tabular}}}

& \textbf{0}         & 0  		& 0      & 0       & 0  	     & 0    & 0.03 & 0  \\
& {\textbf{0.032}}	 & 0.03     & 0.04   & 0       & 0       & 0.03 & 0.05 & 0.03  \\
&{\textbf{0.096}}    & 0.07     & 0.07   & 0.04    & 0       & 0    & 0.09 & 0.04  \\
&{\textbf{0.287}}    & 0.11     & 0.11   & 0.08    & 0.08    & 0.09 & 0.11 & 0.10   \\ \hline 

\multirow{4}{*}{{\begin{tabular}[c]{@{}>{\columncolor{black!5}}c@{}}Unbalanced\\ poorly-connected network \\{\small(15 studies in total)}\end{tabular}}}

& \textbf{0}       & 0         &  0     & 0     & 0     & 0    & 0.06 & 0.06 \\
& {\textbf{0.032}} & 0.06      &  0.08  & 0.06  & 0     & 0.13 & 0.13 & 0.06 \\
& {\textbf{0.096}} & 0.12      &  0.12  & 0.06  & 0.06  & 0.13 & 0.02 & 0.07 \\
& {\textbf{0.287}} & 0.12      &  0.13  & 0.12  & 0.12  & 0.13 & 0.02 & 0.14 \\
\bottomrule
\end{NiceTabular}
}
\label{tab2}

\end{table}

Table \ref{tab2} shows that methods based on posterior predictive $p$-values led to smallest false-positive rates, on average and in unbalanced cases, the rate was slightly higher under the likelihood-based discrepancy compared to the other discrepancy measures. This might be due to the fact that the likelihood contribution of each study is itself affected by the heterogeneity parameter, which in these cases lead to very small $f_i\left(\mathcal{D}_i^\ast\middle|\boldsymbol{\theta}_{0}\right)$. Conversely, $f_i\left(\mathcal{D}_i\middle|\boldsymbol{\theta}_{0}\right)$ can be large, as the values observed values can be quite dispersed. Clearly, this leads to very small $p$-values, likely to be falling below the threshold of outlyingness (see Supplementary material). This seems also in agreement with the findings in \citealt{Zhang2015}. 

\begin{figure}[!h]
\begin{center}
	\includegraphics[scale=0.074]{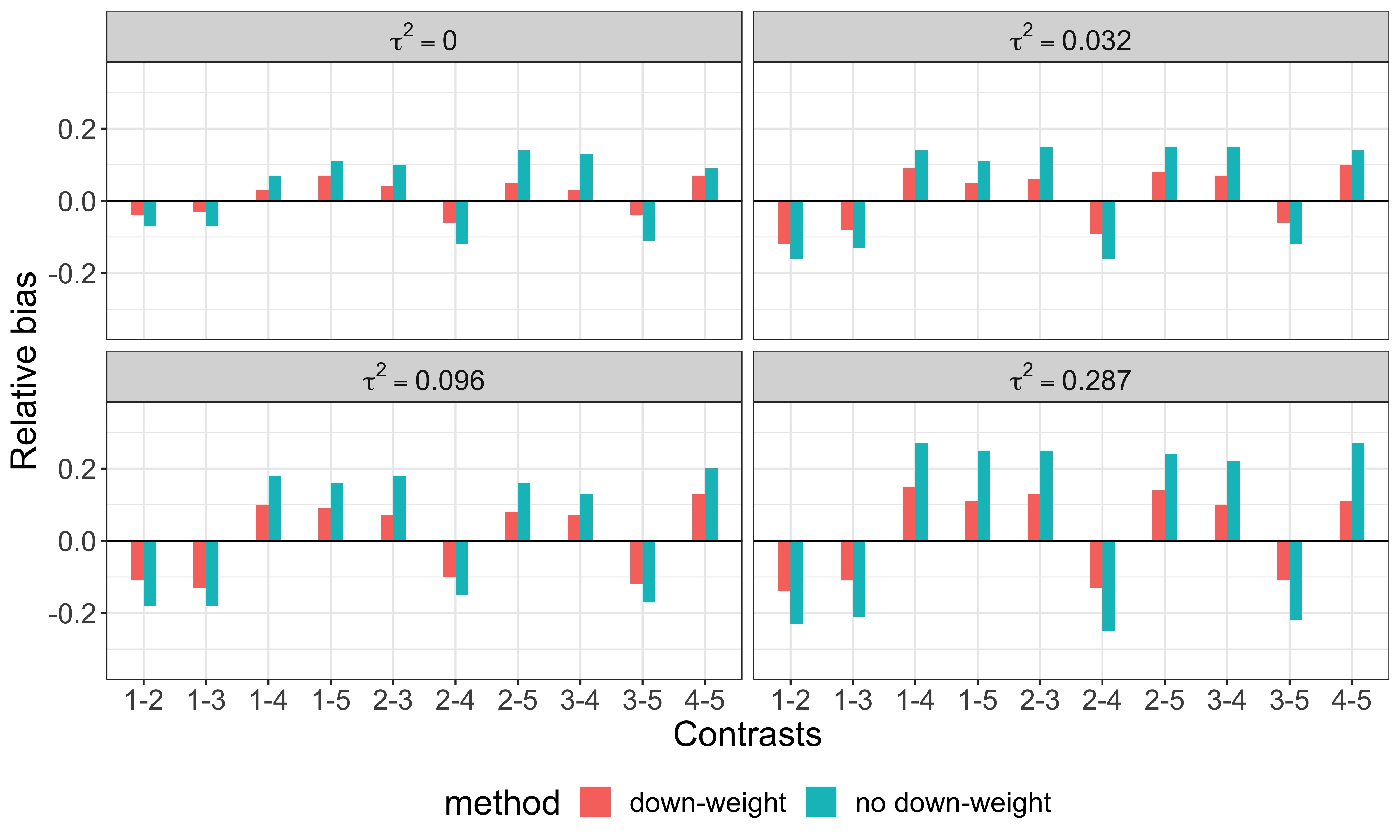}
\end{center}
\caption{Relative bias plot for the network meta-analysis estimates with and without down-weight out of 1000 simulated data sets, at varying heterogeneity, for the case of an unbalanced design with poorly connected network and three induced outliers.}	
\label{fig:fig1}
\end{figure}

Finally, to assess the performance of our down-weighting scheme on each contrast estimate, we computed the estimate relative bias. Figure \ref{fig:fig1} reports the effect of the down-weighting method on the estimate biases at varying heterogeneity, for the unbalanced scenario with poorly connected network and three artificial outliers, which is associated with the highest down-weighting benefit. The contrast estimates which show highest bias refer to the treatment comparisons of the outlying studies and in some cases, of treatment comparisons informed by very few studies. Full results for the other scenarios can be found in the Supplementary material. In all scenarios, down-weighting the suspicious studies is almost always associated with less biased estimates, with magnitude of benefit increasing at larger heterogeneity, in particular for those contrast for which direct evidence is available.

\section{Applications}
\label{sec:appli} 

 In this section we apply our proposed outlier-detection tools to the two motivating networks of interventions previously described in Section \ref{sec:data}, and illustrated in Figure \ref{fig:fig_netplots}. For lung cancer data, we used objective response (ObR) - defined as a complete response or a partial response according to the Response Evaluation Criteria in Solid Tumors \citep{Therasse2000} - while smoking cessation data report the number of individuals who successfully quit smoking after $6$ to $12$ months. In both cases, the odds ratio (OR) was used as a summary measure.

\begin{figure}[p]
\begin{center}
{
      \includegraphics[width=0.99\textwidth,height=0.7\textheight,keepaspectratio]{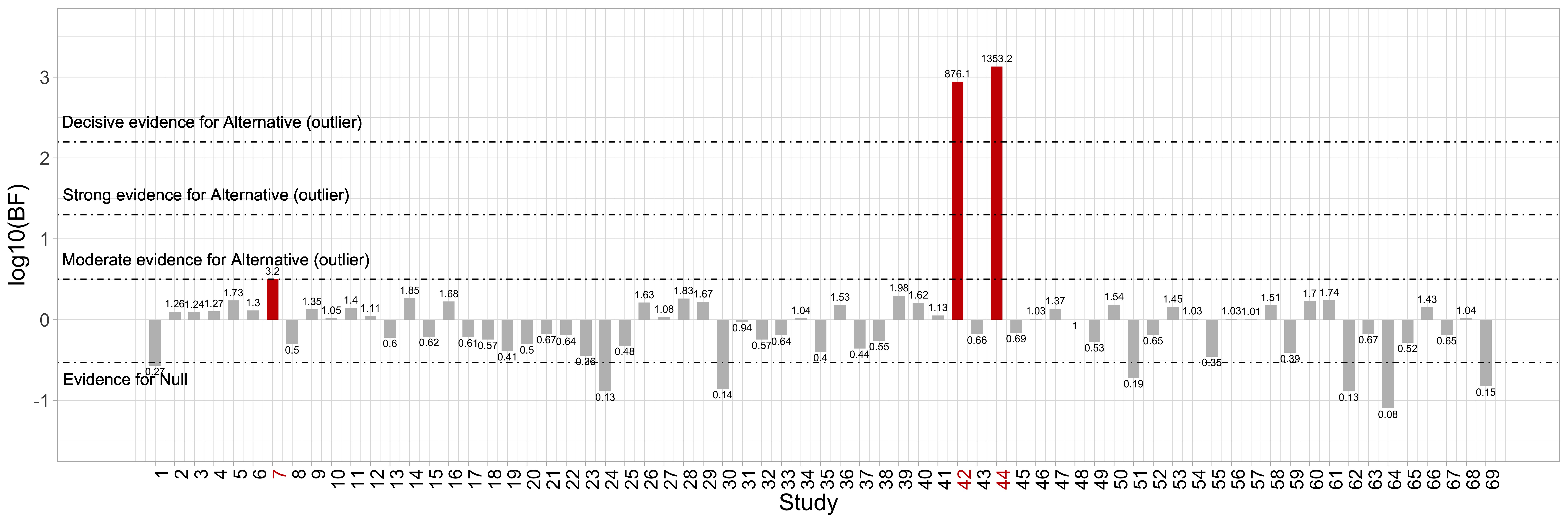} 
   
      \includegraphics[width=0.974\textwidth,height=0.7\textheight,keepaspectratio]{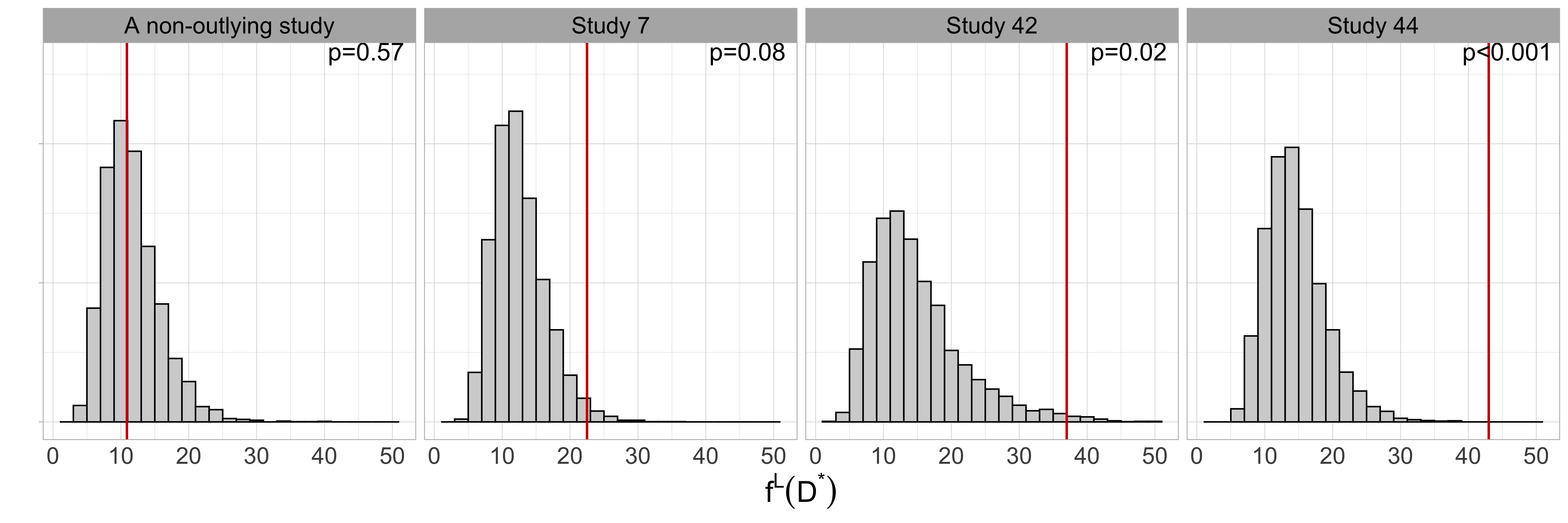}
}%
\vfill 
{
      \includegraphics[width=0.99\textwidth,height=0.7\textheight,keepaspectratio]{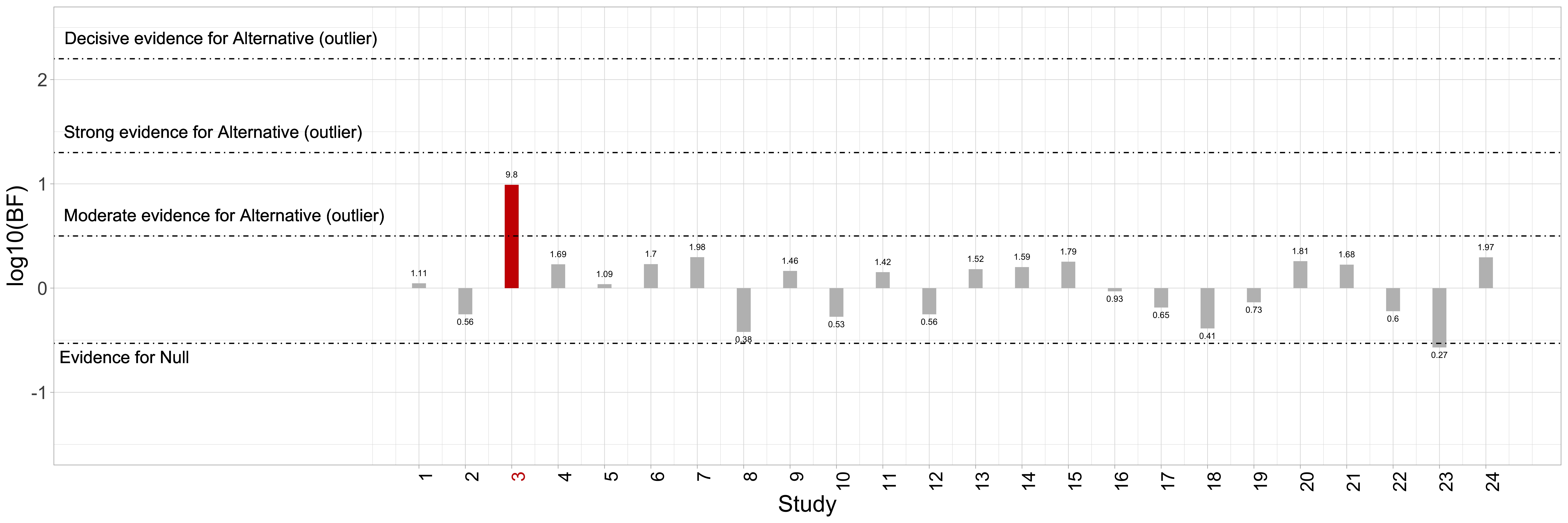}
  
      \includegraphics[width=0.5\textwidth,height=0.7\textheight,keepaspectratio]{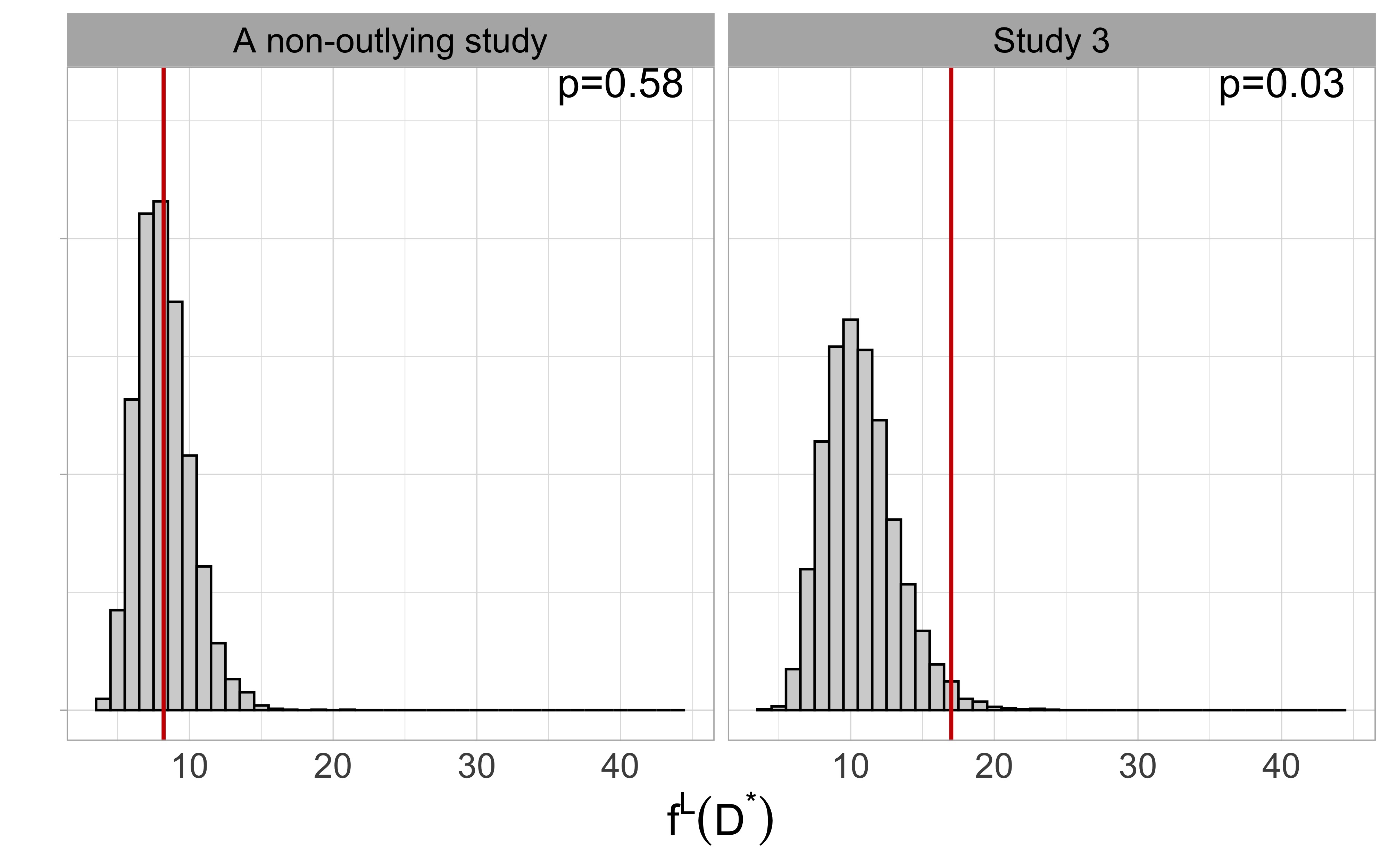}
      }
\end{center}
\caption{Bayes factors with thresholds of evidence (annotated values on the linear scale) and histograms of draws from the posterior predictive distribution for the replicated vs.realised likelihood (vertical line) for the potential outliers identified under the likelihood-based discrepancy measure $f$, alongside a randomly chosen non-outlying study used as comparison (with annotated posterior predictive $p$-values). The two upper plots correspond to lung cancer data, and the lower two plots to smoking cessation data.}
\label{fig:bf_pppc}
\end{figure}

In Figure \ref{fig:bf_pppc}, we report the estimated Bayes factors for each study and the posterior predictive distributions for the detected outliers under the likelihood-based discrepancy. Similar results were achieved under the Stahel-Donoho outlyingness (SDO) discrepancy and can be found in the Supplementary material. In both data sets, we used $50000$ iterations for two MCMC chains and a burn-in period of $10000$ samples. {\color{black} Vague normal priors, $N(0,1000)$, were used for the fixed effect and for each basic parameter and location-shift parameter, and a vague uniform distribution, $U(0,5)$, was used for the heterogeneity $\tau^2$.} Our diagnostic tools detected three potential outliers in the NSCLC network. Here, study $44$ and $42$ were associated with large Bayes factors and relatively small predictive $p$-values, in support of a strong or decisive evidence in favour of outlyingness, while study $7$ is associated with a relatively low Bayes factor and high $p$-value, suggesting a low evidence of outlyingness. In the smoking cessation network, one potential outlier was identified (study 3), associated with moderate Bayes factor and predictive $p$-value. This study was also identified as outlying in \citet{Petropoulou2021}. {\color{black} In the lung cancer network, most included studies have unknown status for epidermal growth factor receptor (EGFR), while study $44$ and study $42$ included Asian patients with respectively wild-type mutation and KRAS (Kirsten Rat Sarcoma Virus) mutation. Compared to the few other included studies with these types of mutations, study $44$ and study $42$ (both comparing Monochemotherapy vs Immunotherapy) have considerably larger proportions of nonsmokers, and these patients are known to vastly differ from smokers in terms of driver mutations and therapy responsiveness (Immunotherapy in particular). }

\begin{figure}[h]
\begin{center}
	\includegraphics[scale=0.082]{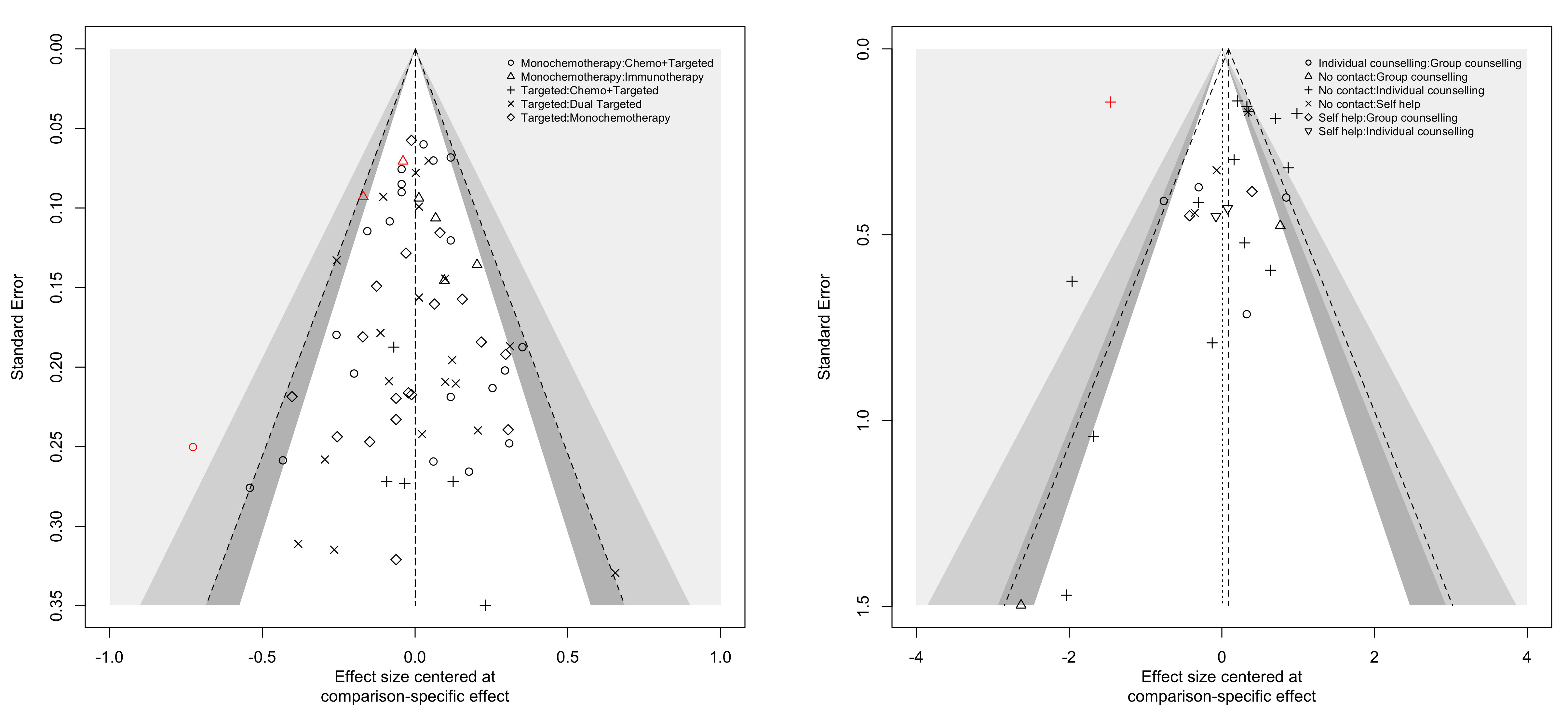}
\end{center}
\caption{Comparison-adjusted funnel plots centered at comparison-specific effect with pseudo confidence intervals at 90\%, 95\% and 99\% approximate confidence levels for the lung cancer data (left panel) and smoking cessation data (right panel). Each comparison-specific effect is plotted against their reversed standard error to further investigate the distribution of the effect sizes. Treatments ordered from oldest to newest in both networks. Studies in red correspond to the potential outliers detected. }	
\label{fig:figfunnel}
\end{figure}

Further, the impact of so-called `small-study effects' was assessed graphically though comparison-adjusted funnel plots, which can in some cases raise additional flags of outlyingness. {\color{black} Here, study $3$ in the smoking cessation data creates an asymmetry in the plot (see Figure \ref{fig:figfunnel})} but interestingly, neither study 42 nor study 44 are identified as suspicious,  supporting the need of sophisticated methods to be used rather than relying on simple visual inspection of funnel plots or standardised residuals. {\color{black} A second stage of analysis was then performed to down-weight these potential outliers, as described in Section \ref{sec:down}. The choices of the beta hyperparameters were made according to the degree of outlyingness of each study. For lung cancer data, a $\mbox{Beta}(3,3)$ - which is centred around 0.5 - was used for study 7 in the lung cancer data and study 3 in the smoking cessation data to reflect the large uncertainty about outlyingness. For study $42$ and $44$, we employed a beta distribution more concentrated in the range $(0, 0.5)$, i.e. $\mbox{Beta}(2,5)$, as we have stronger evidence in favour of outlyingness and so wish to apply a more severe downgrading.  We refer the reader to Figure 1 in Supplementary material for a visual inspection of the chosen beta distributions.}

\begin{figure}[h]
\begin{center}
    \captionsetup{justification=centering}
    \begin{subfigure}{0.45\textwidth}
    \hspace{-1cm}
      \includegraphics[scale=0.097]{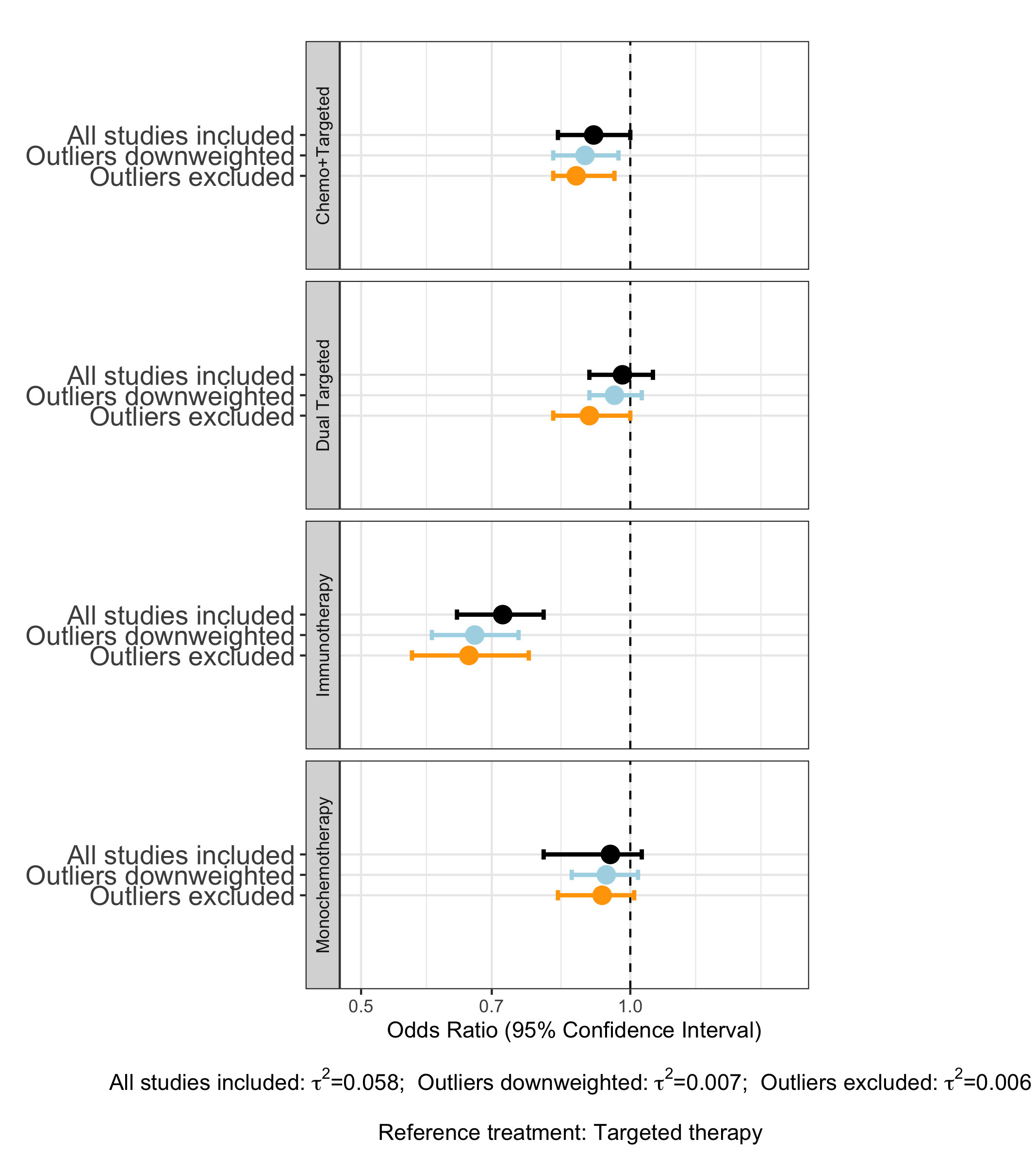}
    \end{subfigure} \hspace{-1.1cm}
    \begin{subfigure}{0.45\textwidth}
      \includegraphics[scale=0.096]{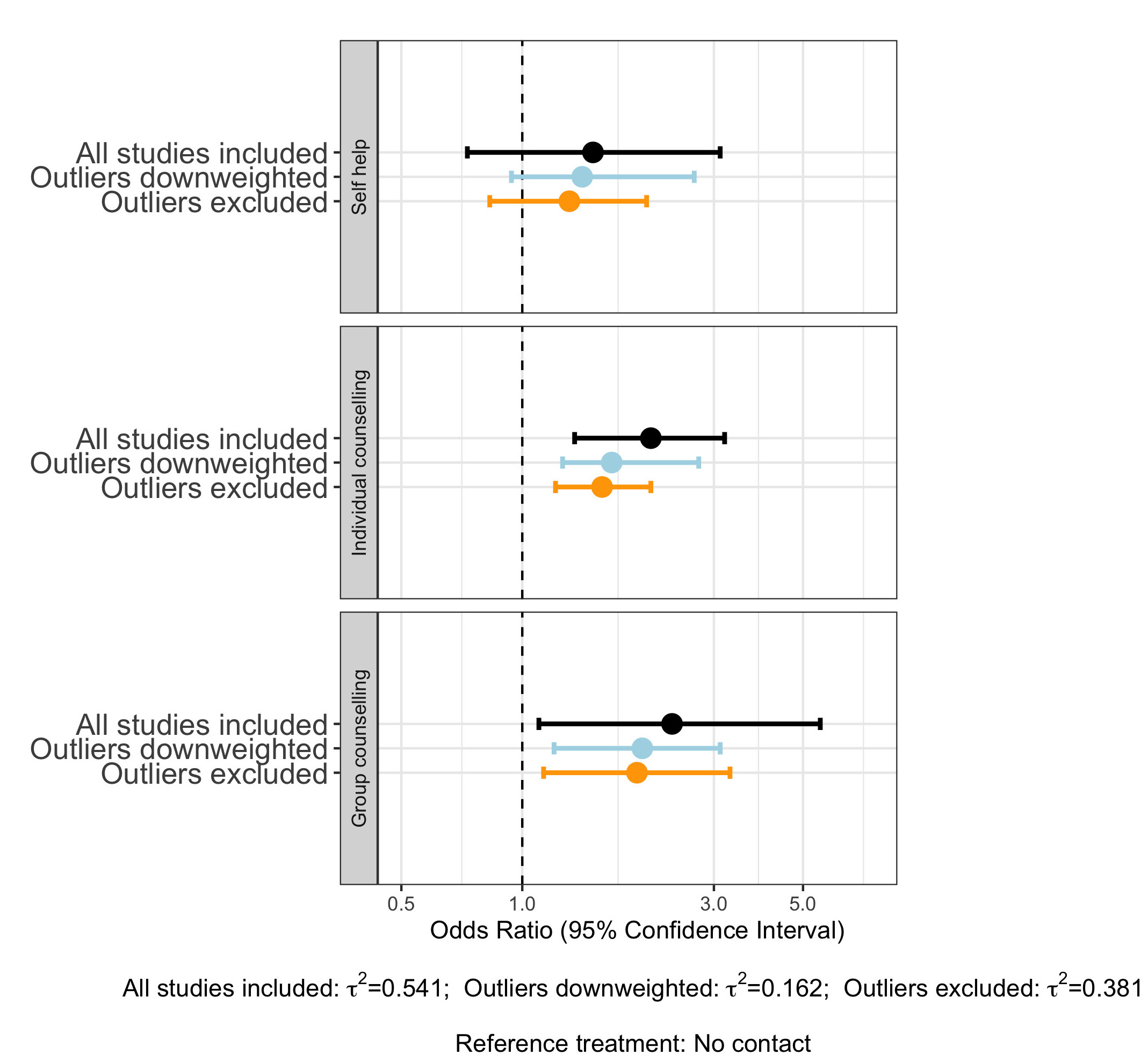}
    \end{subfigure}
\end{center}
\caption{Forest plots of network estimates for all studies included, outliers down-weighted and outliers excluded in the lung cancer data (left) and smoking cessation data (right).} 
\label{fig:figforest}

\end{figure}

Finally, we assessed the robustness of our results comparing the network estimates with and without down-weighting, and when outliers are removed from the network. For lung cancer data, we observe a reduction in the heterogeneity estimates both when the three studies are down-weighed and excluded. Study 44 was associated with the highest contribution matrix percentages (full contribution matrix reported in the Supplementary material). The contribution matrix \citep{Therasse2000} measures how much each direct treatment effect contributes to the effect estimate from network meta-analysis and can support detection of influential studies. However, as shown in Figure \ref{fig:figforest}, moderate changes in the comparative ORs were observed in the overall estimates, where the most significant change is in the effect of Immunotherapy vs. Targeted therapy, which changed from 0.72 (95\%CI: 0.64-0.80) to 0.67 (95\%C I: 0.60-0.75). For the smoking cessation data, the down-weighting of Study 3 (No contact vs. Individual Counselling) markedly reduced the estimated heterogeneity (from $\tau^2  = 0.541$ to $\tau^2 = 0.162$) and thus, the standard error estimates of the ORs became smaller as a whole. In particular, the comparative OR of Individual Counselling vs. No Contact was changed from 2.09 (95\%CI: 1.35-3.19) to 1.67 (95\%CI: 1.26-2.75) with down-weighting and 1.58 (95\%CI: 1.21-2.09) with study exclusion. Here, down-weighting study 3 appears a more conservative choice, as with relatively small networks the exclusion of even a single study can affect significantly the overall estimates.

\section{Discussion}
\label{sec:concl}

In this paper, we have proposed two model-based methods to detect outlying studies in network meta-analysis, leveraging Bayes factors and posterior predictive assessments,  and we have further presented a simple scheme to down-weight the studies detected. 
We have focused on binary data, but the methods can be applied to any type of data. All proposed methodology was tested both on simulated and empirical data.

In simulations, we have identified most of the artificially induced outliers, although both methods fail to some degree to detect outliers with poorly connected networks, with few studies per comparison and mostly, at increasing heterogeneity. This is relatively expected as outliers may in fact cause heterogeneity to be overestimated and in turn affect procedures to detect them, especially when there is not enough information available in the network. Posterior predictive $p$-values achieved the best detection power in comparison with Bayes factors, under both the likelihood-based and SDO-based discrepancy. The forward search (FS) algorithm and, in several scenarios, the cross-validation conditional predictive ordinates (CPO) computed via INLA were also outperformed.Likewise posterior predictive $p$-values, CPO is a Bayesian diagnostic tool based on predictive densities but does not always discriminate well outliers from influential data, as points with high leverage may have small CPOs, independently of whether or not they are outliers. This suggests that the use of posterior predictive $p$-values with discrepancy measures able to capture extreme deviations are essential to improve the detection performance within network meta-analyses, as also pointed out by \citealt{Zhang2015}. When we used our approaches in an network meta-analysis of 112 randomised controlled trials comparing second-line treatments for advanced NSCLC, we identified one clear and two potential outliers corresponding to very large and moderate Bayes factors and posterior predictive $p$-values. In the well-known smoking cessation data, we identified one potential outlier, with a moderate Bayes factor and $p$-value. The down-weighting scheme yielded a significant reduction in the bias of the relative effect sizes estimates in simulations, suggesting the scheme to be effective; which was also confirmed on real data by an overall reduction in heterogeneity and more precise confidence intervals of the network meta-analysis estimates. In the smoking cessation data, it also led to a clear reduction in the contrast estimate related to the outlying study, suggesting it to be also influential.

With both simulated and real data, the different detection methods were not always in full agreement, confirming that it is good practice to jointly assess more than one measure when searching for outliers. Indeed, our proposed tools should not be seen as competing alternatives, but rather as complementing each other and should ideally be used in combination. This is because they capture different aspects of the modelling mechanism: while Bayes factors can be used to compare models (in our case a standard model versus an outlier mean-shift model), posterior predictive $p$-values can only assess discrepancy between the observed data and some assumed model. A reason in support of the Bayes factor is that it is based on weighing the alternative models by the posterior evidence in favour of each of them and thus can also measures evidence in favour of the null hypothesis. Similarly, posterior predictive $p$-values can represent powerful tools for assessing outliers in a Bayesian fashion, but require careful choice of the discrepancy measure, that should always be chosen according to the scientific context and question of interest. 

Our proposed tools present also limitations. For example, Bayes factors are known to be dependent on the choice of the prior distributions and thus caution is needed, especially when informative priors are used into the network meta-analysis model. Moreover, our Bayes factor test depends on how the alternative model is defined. In this paper, the outlier model was constructed as a mean-shift model, but more sophisticated approaches, for example incorporating both a shift in mean and in variance, could be considered. Under certain circumstances, this would aid to account into the model for sample size or related phenomena such as small-study effects. Overall, the method searches for one outlier at the time, making it subject to well-known masking problems (e.g. when a cluster of outliers shift the model parameters to a degree that makes these observations not being identifiable as outliers). Accounting for multiple outliers simultaneously is a topic of further research which would require external knowledge about the groups of studies to be tested to achieve computational feasibility. 
The posterior predictive $p$-value assessment could alternatively be carried out in a cross-validatory leave-one-out setting but it would become computationally intensive, which can be problematic when the network is large \citep{Marshall2003}. Regarding the discrepancy measures chosen, one limitation of the Stahel-Donoho measure is that it implicitly assumes the non-outlier data to be symmetrically distributed and thus it may fail to detect asymmetry in very skewed data. Other choices can include the skewness-adjusted outlyingness (AO) measure \citep{Brys}. Assessment of inconsistency was out of scope in this paper, but we should acknowledge that outlying studies can also be the primary source of inconsistency; in which case differentiating between outlyingness and inconsistency would be difficult: as with heterogeneity, outliers may contribute significantly to an increased inconsistency in the network whilst at the same time affecting the inconsistency checking procedures.
 
In conclusion, our methods have shown encouraging outlier detection results, but we advise that they should always be used in conjunction with clinical expertise and judgement. Looking at future work, we are interested in extending the methodology in a multiple outcome framework  \citep{Efthimiou2015}, to see whether a study has an outlying behaviour in all the reported outcomes. Clearly, this would allow to draw more precise conclusions about the outlyingness of each study in the network. Finally, our simple down-weighting scheme could be refined to allow automatic down-weight of the outliers, rather than specifying the down-weighting factors for outlying studies only at a second stage of analysis. Again, expert information could be used for constructing more appropriate down-weighting factors and further sensitivity analyses may be added to compare the choice of different prior weights.  
The source code for the proposed methods, which we further plan to incorporate into an R package to facilitate broader usage, is freely available at \url{https://github.com/silviametelli/Bayes-NMA-outlier-detection}.





\end{document}